\documentclass[prb,aps,twocolumn,showpacs,eqsecnum,superscriptaddress]{revtex4-1}
\usepackage{amssymb}
\usepackage{amsmath}
\usepackage{graphicx}
\usepackage{wasysym}
\usepackage{mathtools}

\newcommand{\be}{\begin{eqnarray}}
\newcommand{\ee}{\end{eqnarray}}
\newcommand{\bbm}{\begin{bmatrix}}
\newcommand{\ebm}{\end{bmatrix}}
\newcommand{\bpm}{\begin{pmatrix}}
\newcommand{\epm}{\end{pmatrix}}
\renewcommand{\v}[1]{{\bf #1}}

\usepackage{color}

\begin{document}
\title[]{Unified bulk-boundary correspondence for band insulators}

\author{Jun-Won \surname{Rhim}}
\affiliation{Max-Planck-Institut f{\"u}r Physik komplexer Systeme, 01187 Dresden, Germany}
\affiliation{Center for Correlated Electron Systems, Institute for Basic Science (IBS), Seoul 08826, Korea}
\affiliation{Department of Physics and Astronomy, Seoul National University, Seoul 08826, Korea}
\author{Jens H. \surname{Bardarson}}
\affiliation{Max-Planck-Institut f{\"u}r Physik komplexer Systeme, 01187 Dresden, Germany}
\affiliation{Department of Physics, KTH Royal Institute of Technology, Stockholm, SE-106 91 Sweden}
\author{Robert-Jan \surname{Slager}}
\affiliation{Max-Planck-Institut f{\"u}r Physik komplexer Systeme, 01187 Dresden, Germany}

\begin{abstract}
The bulk-boundary correspondence, a topic of intensive research interest over the past decades, 
is one of the quintessential ideas in the physics of topological quantum matter. 
Nevertheless, it has not been proven in all generality and has in certain scenarios even been shown to fail, depending on the boundary profiles of the terminated system.
Here, we introduce bulk numbers that capture the exact number of in-gap modes, without any such subtleties in one spatial dimension. 
Similarly, based on these 1D bulk numbers, we define a new 2D winding number, which we call the pole winding number, that specifies the number of robust metallic surface bands in the gap as well as their topological character.
The underlying general methodology relies on a simple continuous extrapolation from the bulk to the boundary, while tracking the evolution of Green's function's poles in the vicinity of the bulk band edges.
As a main result we find that all the obtained numbers can be applied to the known insulating phases in a unified manner regardless of the specific symmetries. Additionally, from a computational point of view, these numbers can be effectively evaluated  without any gauge fixing problems.
In particular, we directly apply our bulk-boundary correspondence construction to various systems, including 1D examples without a traditional bulk-boundary correspondence, and predict the existence of boundary modes on various experimentally studied graphene edges, such as open boundaries and grain boundaries. 
Finally, we sketch the 3D generalization of the pole winding number by in the context of topological insulators.
\end{abstract}


\maketitle

%%%%%%%%%%%%%%%%%%%%%%%%%%%%%%%%%%%%%%%%%%%%%%%%%%%%%%%%%%%%%%%%%%%%%%%%%%%%%%%%%%%%%%
%%%%%%%%%%%%%%%%%%%%%%%%%%%%%%%%%%%%%%%%%%%%%%%%%%%%%%%%%%%%%%%%%%%%%%%%%%%%%%%%%%%%%%

\section{Introduction}

Topological order has been an active theme in condensed matter physics over the past decades. 
With the discovery\cite{Klitzing1980} of the quantum Hall effect (QHE), in particular, it became apparent that topological concepts are needed for the description of certain quantum orders\cite{Thouless1982,Berry1984,Haldane1988} in addition to the usual symmetry-based classification schemes. 
The according QHE invariant then plays a role analogous to conventional order parameters and corresponds to a physical observable, being the quantized Hall conductance\cite{Thouless1982}. 
This integer is in turn related to the number of protected chiral edge states by virtue of the system being a free-electron insulator in the bulk. 
More recently, topological considerations were revived in the context of band structures \cite{Hasan2010,Qi2011}. 
That is, it was found that the concepts of symmetry and topology can be combined, resulting in (nearly) free fermions states that feature a topological invariant as a result of the presence of a symmetry\cite{Fu2007a,Moore2007,Schnyder2008,Kitaev2009,Fu2011,Slager2013,Chen2013,Chiu2014,Chiu2016}. 
Following the prediction and experimental discovery of many time-reversal protected Z$_{2}$ topological band insulators\cite{Bernevig2006,Kane2005,Fu2006,Fu2007,Fu2008,Konig2007,Hsieh2008,Xia2009,Zhang2009}, the active investigation of such symmetry protected topologically ordered states and their associated physical consequences has in fact also been driven by the identification of many actual material candidates.

While the impact of the topological entity can be traced from a bulk perspective\cite{Qi2008scs,Ran2008scs,Juricic2012,Slager2014,Tteo2010,imuraprb2011,Alexandradinata2016,Slager2016}, a highlight of topological order is formed by the presence of signature edge states via a bulk-boundary correspondence (BBC) similar to the QHE case. 
These edge states have both direct experimental and theoretical consequences\cite{Mong2011,Essin2011,Slager2015,Konig2007,Hsieh2008,Xia2009,Zhang2009}.
Indeed, edge states can directly be experimentally verified using ARPES measurements, whereas the halving of the degrees of freedom lies at the basis of new theoretical proposals including the notable possibility of excitations having fractional charges and statistics\cite{Fu2008r}. For example, in a topological insulator, each spatially separated edge hosts a {\it single} Dirac cone.
However, a general relation between the bulk and boundary modes is yet to be established and thus forces one to case-by-case evaluations. 
In case of a topological phase that is, e.g., solely protected by crystalline symmetries, the termination that results in the boundary has to at least respect the protecting symmetry\cite{Rhim2017,Fang2015,Hirayama2017}.
More generally, one can note that, by the incompressible nature of the bulk topology, the bulk system features a robustness that is set by the bulk band gap, whereas the edge states can in principle immediately be gapped  by a symmetry breaking perturbation\cite{Sheng2006,Qi2006,Li2010,Prodan2009}.
Furthermore, in 1D reflection symmetric insulators, the Zak phase's\cite{Zak1989}(or Berry phase's) BBC requires the commensurability between a certain choice of the bulk unit cell and the terminated system, while the finite system should also remain insulating\cite{Rhim2017,Vanderbilt1993b}. 
Since the latter condition cannot be checked from the bulk perspective, the Zak phase misses the complete prediction of the number of in-gap boundary modes. 
This can be exemplified in many specific models including a coupled Su-Schrieffer-Heeger(SSH) model that we will employ later.

Although BBCs were considered as early as the 1930's\cite{Tamm1932,Maue1935,Goodwin1939a,Goodwin1939b,Goodwin1939c,Shockley1939,Davison1970,Kalkstein1971,Lee1981,Zak1984,Zak1985}, we here universally address the role of the BBC. That is, we identify direct measures to predict the appearance of midgap states between two bands in the presence of a general boundary. 
In particular, starting from a construction that can directly be linked to Green's functions, we define simple quantities calculated from bulk wave functions that directly convey the number of modes in the gap of 1D insulators. 
%as well as the topological character of edge states protected by the underlying symmetry, i.e TRS for a $Z_2$ topological insulator.
%
As a next step, we then lift these ideas to 2D, culminating in the concept of the pole winding number which is completely distinguished from the usual definitions of winding number based on the TKNN number.
Specifically, we find that the trajectories of the poles of the Green's function and their  chiralities relate to the presence of a topological invariant in the bulk.
Although this winding number is obtained from bulk wave functions, it predicts the number of in-gap surface bands. 
Moreover, evaluating the winding of the poles in detail also discerns whether these bands have a topological status in the sense that it conveys whether they connect the valence and conduction bands.
The resulting number can be evaluated in a unified way regardless of underlying symmetries of the system. 
After elucidating all these notions with specific examples, including well-kown models such as the Kane-Mele model\cite{Kane2005} and experimentally studied graphene grain boundaries\cite{Lahiri2010,Kim2011,Huang2011,Tsen2012,Biro2013,Phillips2015}, we finally also sketch the applications of our ideas to 3D.

The rest of this paper is organized as follows. 
In Section \ref{sec:z_number} we set the stage and introduce some essential concepts as well as the underlying idea of why an evaluation of the poles of the Green's functions relates to the topological bulk invariant and therefore sheds light on the BBC of the system.
This then leads to the identification of robust numbers that convey the number of in-gap modes and their topological character.
Subsequently, in Section \ref{sec::z2}, we then link the previous notions to systems having a  topological Z$_{2}$ classification.
Then, we apply our general machinery to numerous specific examples in Section \ref{sec::examples} to elucidate the more formal preceding sections. 
In Section \ref{sec::pole} we show that this strategy naturally leads to the pole winding number. 
This number can similarly  be used to predict the number of edge states as well as their topological character in 2D, whereas the generalization to 3D is implicitly evident. 

\begin{figure}
	\begin{center}
		\includegraphics[width=0.85\columnwidth]{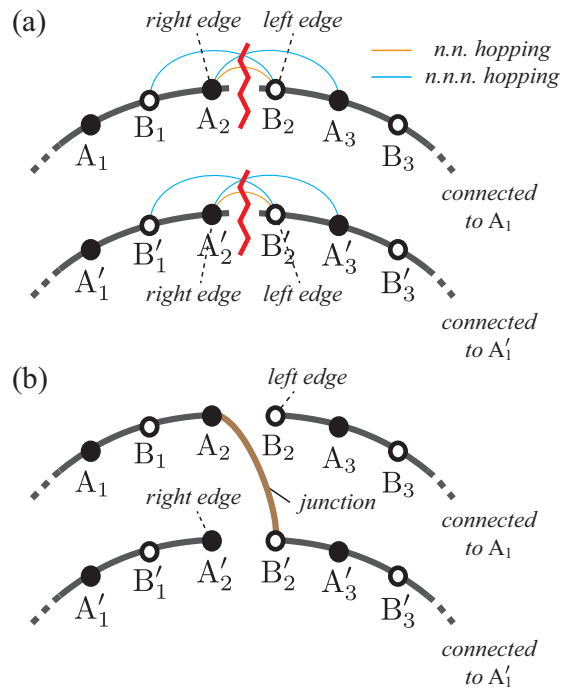}
	\end{center}
	\caption{ Schematic description of the use of local potentials for making open boundaries and junctions by considering tight binding models with nearest neighbour (\textit{n.n}) and next nearest neighbour (\textit{n.n.n.}) hopping processes. (a) We prepare two bulk systems represented by two ring geometries. To make open boundaries (red zigzag line) between A$_2$ and B$_2$ in the upper chain, we add a local potential that cancels all the hopping processes (yellow and blue curves) crossing this red line. The same applies to the lower chain. (b) Given two chains with open boundaries, we add another local potential that makes a connection (brown line) between A$_2$ and B$_2^\prime$ which completes the junction between the upper and lower chains.
	}
	\label{fig:local_pot}
\end{figure}

\section{Obtaining the number of in-gap modes from bulk properties}\label{sec:z_number}
We first explain the method of obtaining the number of in-gap modes for the case of one spatial dimension and consider a general translationally invariant system with an arbitrary number of bands.
With periodic boundary conditions, such systems can be described on a ring geometry with $N$ unit cells, which we refer to as the \textit{bulk}.
Various terminations of the bulk, as well as junctions between two different bulks, are generally realized by adding a local operator $\mathcal{V}_b$ to the bulk Hamiltonian $\mathcal{H}_0$. 
Consequently, we study a system described by a Hamiltonian 
\begin{equation}\label{eq::central}
\mathcal{H} = \mathcal{H}_0 + \mathcal{V}_b.
\end{equation}
For instance, for an open boundary $\mathcal{V}_b$ consists of hopping terms that cancel all the hopping terms of the bulk Hamiltonian that cross the boundary between two neighboring unit cells, as illustrated in Fig.~\ref{fig:local_pot}(a).
Similarly, for a junction between two distinct bulks, we can modify two independent ring geometries into two finite systems with open boundaries, following the above prescription, and then apply additional hopping terms to $\mathcal{V}_b$ that connect the two terminated systems, as depicted in Fig.~\ref{fig:local_pot}(b).
Higher dimensional systems can similarly be studied by constructing an effective 1D Hamiltonian for each fixed transverse momenta. 

Next we lift this construction to a parameter family of Hamiltonians 
\be
\mathcal{H}_{\beta} = \mathcal{H}_0 + \beta\mathcal{V}_b
\ee
where $\beta$, varying from 0 to 1, extrapolates between the periodic bulk and the terminated system.
To obtain a bulk criteria for the existence of in-gap localized modes of the system with an edge, $\mathcal{H}_{\beta=1}$, we simply count the net number of states that are transferred from the bulk band continuum into the band gap. 
Note that we may presume those in-gap modes to be localized as they result from a local potential $\mathcal{V}_b$ that cannot affect the bulk wave functions far away from the local region.

\begin{figure}
	\begin{center}
		\includegraphics[width=\columnwidth]{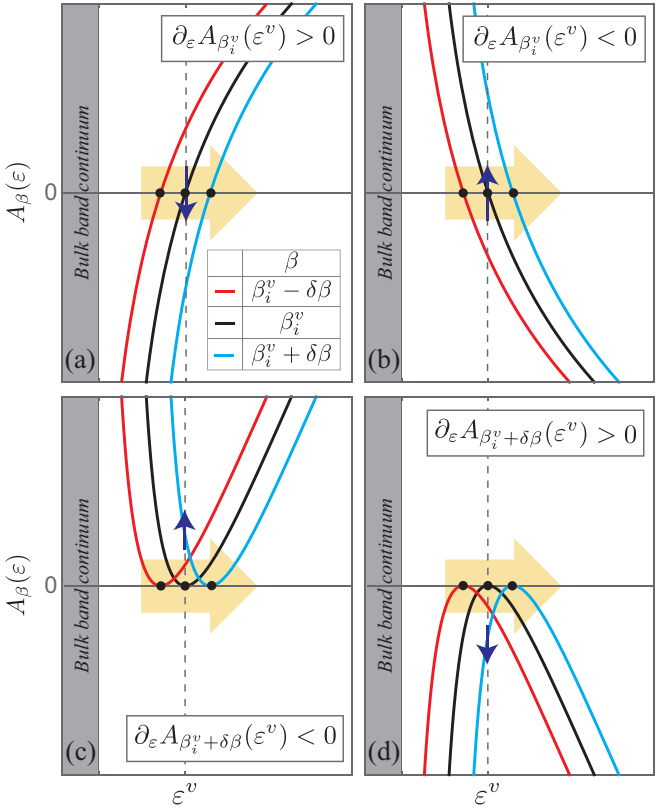}
	\end{center}
	\caption{Plot of $A_\beta(\varepsilon)$ as a function of $\varepsilon$ for $\beta= \beta^v_{i}-\delta\beta$, $\beta^v_{i}$ and $\beta^v_{i}+\delta\beta$ in the vicinity of the valence band portal ($\varepsilon^v$) when some modes (dots at $A_\beta(\varepsilon)=0$) are coming into the gap from the valence band continuum through the valence band portal (dashed lines). 
	The horizontal yellow arrows denote that the pole enters the valence band portal for increasing $\beta$.
	Blue vertical arrows indicate the $\beta$-derivative of $A_{\beta}(\varepsilon)$ at the valence band portal: the upward (downward) arrow for the positive (negative) slope.
	One mode comes into the gap with (a) $\partial_\varepsilon A_{\beta}(\varepsilon^v)> 0 $ and $\partial_\beta A_{\beta}(\varepsilon^v)  < 0$, and (b) with $\partial_\varepsilon A_{\beta}(\varepsilon^v) < 0$ and $\partial_\beta A_{\beta}(\varepsilon^v)  > 0$ at $\beta = \beta^{v}_{i}$. 
	Two degenerate modes are introduced into the gap (c) with $\partial_\varepsilon A_{\beta}(\varepsilon^v) < 0$ and $\partial_\beta A_{\beta}(\varepsilon^v)  > 0$, and (d) with $\partial_\varepsilon A_{\beta}(\varepsilon^v) > 0$ and $\partial_\beta A_{\beta}(\varepsilon^v)  < 0$ at $\beta = \beta^{v}_{i}+\delta\beta$.}
	\label{fig:A_plot}
\end{figure}

The number of modes $M$ in the gap between the valence and conduction band edges, is generally given by 
\begin{equation} \label{eq:number of surface modes}
M = -\frac{1}{\pi}\mathrm{Im}\int^{\varepsilon^c}_{\varepsilon^v} d\varepsilon~ \mathrm{Tr}\mathcal{G}_1(\varepsilon),
 \end{equation}
where $\mathcal{G}_\beta(\varepsilon) = (\varepsilon - \mathcal{H}_\beta +i\eta)^{-1}$ is the retarded Green's function for $\mathcal{H}_\beta$.
From now on, we refer to the retarded Green's function simply as the Green's function.
Here, $\varepsilon^c$ and $\varepsilon^v$ are equipotentials infinitesimally shifted from the conduction and valence band edges into the gap, called the conduction and valence band \textit{portals}.
The infinitesimal shift allows us to assume there are no boundary modes between the portal and the bulk continuum.
Our strategy is to evaluate the integral in Eq.~\eqref{eq:number of surface modes}, assuming that the eigenenergies $\epsilon_{n,k}$ and eigenfunctions $|n,k\rangle$ of the bulk Hamiltonian $\mathcal{H}_0$ are known, by studying the behavior of poles of $\mathcal{G}_\beta(\varepsilon)$ as a function of $\beta$.

The evaluation of $M$ is done by specifying the poles of $\mathcal{G}_1(\varepsilon)$ in the gap.
This is equivalent to the number of roots of $A_{\beta=1}(\varepsilon)$, where
\begin{align}
A_\beta(\varepsilon) = \mathrm{det}\left\{1 -\beta\mathcal{G}_0(\varepsilon)\mathcal{V}_b\right\}\equiv\mathrm{det}\mathcal{A}_\beta(\varepsilon),
\end{align}
since $\mathcal{G}_\beta(\varepsilon) = [1-\beta \mathcal{G}_0(\varepsilon)\mathcal{V}_b]^{-1}\mathcal{G}_0(\varepsilon)$ and $\mathcal{G}_0(\varepsilon)$ has no poles in the gap.
We refer to $\mathcal{A}_\beta(\varepsilon)$ and $A_\beta(\varepsilon)$ as the \textit{pole-matrix} and the \textit{pole-determinant} in the remainder.
The bulk eigenenergies and eigenfunctions are employed in such a way that $\mathcal{G}_0(\varepsilon)\mathcal{V}_b = \sum_{n,k}(\varepsilon - \epsilon_{n,k})^{-1}|n,k\rangle\langle n,k|\mathcal{V}_b$.

Some remarks about the above are in order. (i) Although $\mathcal{A}_\beta(\varepsilon)$ is not Hermitian, $A_\beta = \mathrm{det}(\mathcal{V}^{-1}_b + \beta\mathcal{G}_0)\mathrm{det}\mathcal{V}_b$ is real-valued because both $\mathcal{V}^{-1}_b + \beta\mathcal{G}_0$ and $\mathcal{V}_b$ are Hermitian in the gap. 
(ii) If the rank of $\mathcal{V}_b$ is $N_\mathcal{V}$, one can find a $N_\mathcal{V}\times N_\mathcal{V}$ matrix representation $\mathcal{A}_\beta(\varepsilon)$ of the operator $[1 -\beta\mathcal{G}_0(\varepsilon)\mathcal{V}_b]$ that determines $A_\beta(\varepsilon)$.
For example, for an open boundary between sites $i=1$ and $i=2$ of a 1D tight-binding chain with Hamiltonian $\mathcal{H} = t\sum_i c_i^\dag c_i$, $\mathcal{V}_b = -t(c_1^\dag c_2 + c_2^\dag c_1)$ with rank 2, and the $2\times 2$ pole-matrix is obtained from the basis vectors $|1\rangle = c^\dag_1|0\rangle$ and $| 2 \rangle = c^\dag_2|0\rangle$ by evaluating the matrix elements $\mathcal{A}_\beta(\varepsilon)|_{i_1,i_2} = \langle i_1 |(1 -\beta\mathcal{G}_0(\varepsilon)\mathcal{V}_b) | i_2 \rangle $ where $i_n \in \{1,2\}$. 
(iii) For a fixed $\varepsilon$, $A_\beta(\varepsilon)$ is a real polynomial of $\beta$ of order $N_\mathcal{V}$ whose coefficients are determined from the $\beta$-independent $\mathcal{G}_0(\varepsilon)\mathcal{V}_b$.
(iv) $A_\beta(\varepsilon)$ is a non-singular function because $\mathcal{G}_0(\varepsilon)$ has no singular points in the gap, and $\mathcal{V}_b$ is independent of $\varepsilon$ and $\beta$.
Therefore, the pole-determinant is a smooth function of $\varepsilon$ and $\beta$ given that band dispersions are smooth.

Using the $\beta$-dependence of the pole-determinant at band portals in the above, we obtain a simple expression for $M$ as follows.
Starting from the bulk ($\beta =0$), some modes are pulled out of the bulk band continuum, and then pass through the portal $\varepsilon = \varepsilon^\alpha$ inward or outward with increasing $\beta$, where $\alpha=c$ and $\alpha=v$ represent the conduction and valence bands, respectively.
Let us first consider the pole-determinant at the valence band portal.
When $\beta =0$, the pole-determinant is unity over the whole gap since we assumed there are no in-gap modes before turning on $\mathcal{V}_b$.
We assume that the polynomial $A_\beta(\varepsilon^v)$ of $\beta$ has $l^v$ distinct roots between $\beta=0$ and $\beta=1$, which we denote $\beta_i$.
If one mode from the bulk band continuum comes into the gap for increasing $\beta$ at $\beta = \beta^v_{i}$, it passes through $\varepsilon = \varepsilon^v$  as illustrated in Fig.~\ref{fig:A_plot} (a) and (b).
In this case, the signs of $\partial_\beta A_{\beta^{v}_{i}}(\varepsilon^v)$ and $\partial_\varepsilon A_{\beta^{v}_{i}}(\varepsilon^v)$ are opposite. 
That is, the product $[\partial_\beta A_{\beta^{v}_{i}}(\varepsilon^v)][\partial_\varepsilon A_{\beta^{v}_{i}}(\varepsilon^v)]$ is negative.
In contrast, if a mode that already resides in the gap moves out, merging eventually into the valence band continuum, we observe that $[\partial_\beta A_{\beta^{v}_{i}}(\varepsilon^v) ][\partial_\varepsilon A_{\beta^{v}_{i}}(\varepsilon^v)]$ is positive.
Similarly, multiple degenerate modes may exit or enter the gap simultaneously due to some symmetry.
One can verify that their entrance into the gap is also signaled by negative $[\partial_\beta A_{\beta^{v}_{i}}(\varepsilon^v)] [\partial_\varepsilon A_{\beta^{v}_{i}}(\varepsilon^v)]$, and their exit from the gap by a positive product of derivatives;
however, in this case, we need to replace $\beta^v_{i}$ with $\beta^v_{i} + \delta\beta$ with $\delta\beta$ positive, because the first derivatives $\partial_\beta A_{\beta^{v}_{i}}(\varepsilon^v)$ and $\partial_\varepsilon A_{\beta^{v}_{i}}(\varepsilon^v)$ vanish for multiple roots of $A_\beta(\varepsilon)$. 
In Fig.~\ref{fig:A_plot} (c) and (d), we illustrate this schematically for a doubly degenerate case.
The degeneracy at $\beta= \beta^\alpha_{i}$, denoted by an integer $p^v_{i}$, is manifested by $A_{\beta^v_i}(\varepsilon) \sim (\varepsilon - \varepsilon^v)^{p^v_{i}}$ from the definition of the Green's function.  
On the other hand, one can also note that $A_{\beta}(\varepsilon^v) \sim (\beta - \beta^v_i)^{p^v_{i}}$ for fixed $\varepsilon$.
This can be understood as follows:
If $A_\beta(\varepsilon^v) \sim (\beta - \beta_i^v)^q$ where $q$ is an integer, we can always find a perturbation to $\mathcal{V}_b$ which slightly deforms the pole-matrix into $A_\beta(\varepsilon^v) \sim (\beta - \beta_i^v-\delta\beta_1)\cdots(\beta - \beta_i^v-\delta\beta_q)$ with $\delta\beta_i \neq \delta\beta_j$ for $i\neq j$.
This means, after adding the perturbation, $q$ degenerate poles come into the gap through the portal at different $\beta$'s in turn, and the maximum number of different entrances $q$ should be the same as the $p^v_i$.

Finally, since everything is the opposite at the conduction band portal ($\varepsilon = \varepsilon^c$), the net number of states moving into the gap though the portal $\varepsilon = \varepsilon^\alpha$ is given by
\begin{align}
M^\alpha = \sum_{i=1}^{l^\alpha} m^\alpha p^\alpha_{i} \mathrm{sgn}\left\{ \left[ \partial_\beta A_{\beta^{\alpha}_{i}}(\varepsilon^\alpha) \right]\left[\partial_\varepsilon A_{\beta^{\alpha}_{i}}(\varepsilon^\alpha) \right] \right\} \label{eq:Z_number_cv}
\end{align}
which leads to
\begin{align}
M = M^c + M^v  \label{eq:Z_number}
\end{align}
as the total number of in-gap modes at $\beta=1$.
Here, $m^{c(v)}=-1(1)$ reflects the opposite behavior of the pole-determinant at the conduction and valence band portals.
Note that, due to the property (iii), analyzing $A_\beta(\varepsilon)$ as a function of $\beta$ comes with negligible additional numerical costs once one obtains the matrix $\mathcal{G}_0(\varepsilon)\mathcal{V}_b$ at those portals.
Also, note that the number of in-gap modes introduced by $\mathcal{V}_b$ is determined only by the properties of $A_\beta(\varepsilon)$ near the conduction and valence band edges, instead of scanning the whole gap.
The formula for $M$ can therefore be applied to any gap in the system by simply changing the chemical potential at which it is evaluated.

Note that we can consider the above scheme as a bulk-boundary correspondence because we predict the number of boundary modes generated by the termination $\mathcal{V}_b$ from the bulk eigenfunctions which are used for the evaluation of the pole-determinant.
While we never use the eigenfunctions of the terminated system, the profile of the boundary under consideration is included in the local potential $\mathcal{V}_b$.
This enables us to predict correct number of boundary modes even in the cases where the validity of the Zak phase's BBC depends on certain conditions for the edge profiles such as the conservation of reflection symmetry and commensurability with the bulk unit cell\cite{Rhim2017,Fang2015}.

A further general consequence of the above discussion is that the maximum number of in-gap modes induced by $\mathcal{V}_b$ is $2N_\mathcal{V}$.
Since the number of in-gap modes is the net amount of incoming poles through both portals, if all the roots $\beta^\alpha_{i}$ of $A_\beta(\varepsilon^v)=0$ and $A_\beta(\varepsilon^c)=0$ are of incoming character, the number of in-gap modes is equal to the number of roots $\beta^\alpha_{i}$ in $0\leq \beta \leq1$.
The maximum number of roots is reached when all the roots of two polynomials of $\beta$, $A_\beta(\varepsilon^v)$ and $A_\beta(\varepsilon^c)$, are real-valued, and located between $\beta=0$ and $\beta=1$, which is equal to $2N_\mathcal{V}$ where $N_\mathcal{V}$ is the order of each polynomial.
For example, for a 1D nearest neighbor tight binding model with a single orbital per site, the allowed maximum number of edge modes in each gap is 4 no matter how many basis sites are in the unit cell.

\section{Even-odd prediction of the number of in-gap modes from the bulk \label{sec::z2}}
In the case of 1D insulators, a simpler and more numerically efficient formula that determines whether the number of in-gap modes is even or odd can be obtained.
To start, we note that since a differentiable function has an even (odd) number of roots in an interval if its sign at the two ends 
of the interval are the same (opposite), we have a Z$_{2}$ number of the form
\begin{align}
P = \frac{1}{i\pi}\ln \mathrm{sgn}\left\{A_1(\varepsilon^v)A_1(\varepsilon^c)\right\}.\label{eq:Z2_number}
\end{align}
If $A_1(\varepsilon^\alpha)$ happens to be zero, we shift $\varepsilon^{\alpha}$ closer to the band edge.
In this case, unlike the number $M$ in the previous section, we do not need to know the full $\beta$-dependence of the pole-determinant.

If $\mathcal{H}_\beta$ features a chiral or particle-hole symmetry for all $\beta$, we can obtain a Z$_{2}$ number $P_\mathrm{half}$ for the even-oddness of the number of in-gap modes in each half of the gap around zero energy, $[\varepsilon^v,0]$ or $[0,\varepsilon^c]$.
Since we start from $A_0(\varepsilon^\alpha) = 1$ at $\beta=0$, and the sign of $A_\beta(\varepsilon^\alpha)$ only changes whenever an odd number of modes enter or exit the gap through $\varepsilon = \varepsilon^\alpha$, we have
\begin{align}
P_\mathrm{half} = \frac{1}{i\pi}\ln \mathrm{sgn}\,A_1(\varepsilon^\alpha).\label{eq:Z2_chiral_number}
\end{align}
This number gives more information on the number of in-gap modes than (\ref{eq:Z2_number});
for example, if a chiral symmetric Hamiltonian has two in-gap modes, $P$ cannot confirm the existence of in-gap modes although it  conveys that there are even number of them.
In contrast, $P_\mathrm{half}$ predicts odd number of in-gap modes in each half of the gap in this case which implies the existence of the in-gap modes.
Note that this number cannot be applied to the case where we have odd number of boundary modes at zero energy at $\beta=1$ because one cannot bring odd number of boundary modes into the gap from the bulk with maintaining chiral or particle-hole symmetries through whole $\beta$ from 0 to 1.

One can use the fact $\varepsilon^\alpha$ can be as close as possible to the band edge to analyze the properties of $\mathcal{A}_\beta(\varepsilon^\alpha)$ in more detail.
Let us denote by $\delta\varepsilon$ the distance between the band edge and the nearby band portal. 
We assume there are $N^*$ number of band edges for the valence or conduction band 
at $k^*_l$ ($1\leq l\leq N^*$) with energy $\epsilon^* = \epsilon^\mathrm{max}$ for $\alpha = v$ and $\epsilon^* = \epsilon^\mathrm{min}$ for $\alpha = c$.
We represent the $i$-th eigenvector and eigenvalue of $\mathcal{V}_b$ as $|v_i\rangle$ and $v_i$, where $1\leq i \leq N_\mathcal{V}$.
Then, in the basis of $\mathcal{V}_b$, $\mathcal{A}_\beta(\varepsilon^\alpha)$ can be expressed as
\begin{align}
\mathcal{A}_{\beta}(\varepsilon^\alpha) = \mathcal{I} + \frac{m^\alpha}{\delta\varepsilon} \beta\mathcal{D}^{(1)}_{\alpha} - \beta\mathcal{D}^{(2)}_{\alpha}  \label{eq:pole-matrix}
\end{align}
where $\mathcal{I}$ is the $N_\mathcal{V}\times N_\mathcal{V}$ identity matrix, 
\begin{align}
\mathcal{D}^{(1)}_{\alpha}|_{ij} = \sum_{l} \langle v_i |\alpha,k^*_l\rangle \langle \alpha,k^*_l | \mathcal{V}_b | v_j \rangle \label{eq:D1}
\end{align}
and
\begin{align}
\mathcal{D}^{(2)}_{\alpha}|_{ij} = \sideset{}{'}\sum_{n,k} \frac{\langle v_i | n,k \rangle \langle n,k | \mathcal{V}_b|v_j \rangle }{ \epsilon^* - \epsilon_{n,k}} \label{eq:D2}
\end{align}
Note that we have separated out the sum over the band edge states in $\mathcal{D}^{(1)}_{\alpha}$ by introducing the primed sum in $\mathcal{D}^{(2)}_{\alpha}$ that does not include them.

A general consequence of the above equations is that the Z$_{2}$ numbers $P$ and $P_\mathrm{half}$ 
are determined only from the bulk eigenstates at the bulk band edges if $\mathrm{det}\mathcal{D}^{(1)}_{\alpha} \neq 0$ for the following reason:
while $\mathcal{D}^{(2)}_{\alpha}$ is dominated by the momentum sum around $k^*_l$'s, it is proportional to the number of unit cells $N$ if the dispersion around $k^*_l$ is quadratic and $\langle v_i | \alpha,k^*_l \rangle \langle \alpha,k^*_l | v_j \rangle$ is nonzero.
As a result, if $\mathcal{D}^{(1)}_{\alpha}$ is invertible, that is $\mathrm{det}\mathcal{D}^{(1)}_{\alpha} \neq 0$, we have $A_1(\varepsilon^\alpha) \approx \left[(-1)^{\alpha}/\delta\varepsilon\right]^{N_\mathcal{V}}\mathrm{det}\mathcal{D}^{(1)}_{\alpha}$ for $\delta\varepsilon \ll 1/N^r$ with $r>1$.
On the other hand, if $\mathrm{det}\mathcal{D}^{(1)}_{\alpha} = 0$, which happens when the dimension of the set of band edge state $\{|\alpha,k^*_l\rangle\}$ is smaller than the rank of $\mathcal{V}$($N^*<N_\mathcal{V}$), we need all bulk states to analyze $A_1(\varepsilon^\alpha)$ in general.

Again these formal definitions can be readily understood in particular instances. For the simplest case, when $N_\mathcal{V} = 2$, we have explicit formulae as follows.
First, when $\mathrm{det}\mathcal{D}^{(1)}_{\alpha} = 0$, we have
\begin{align}
A_1(\varepsilon^\alpha) \approx \frac{(-1)^{\alpha}}{\delta\varepsilon} \left( \langle \mathcal{V}_b\rangle^* - \sideset{}{'}\sum_{n,k}\sum_{l} \frac{ d^{n,k}_{\alpha, k^*_l}}{\epsilon^* - \epsilon_{n,k}} \right) \label{eq:2by2_formula_1}
\end{align}
where $\langle \mathcal{V}_b\rangle^* = \sum_{l}\langle \alpha,k^*_l | \mathcal{V}_b | \alpha,k^*_l \rangle$, and $d^{n_1, k_1}_{n_2, k_2} = \prod_{i=1,2}\langle n_i,k_i| \mathcal{V}_b |n_i, k_i \rangle  - |\langle n_1,k_1| \mathcal{V}_b |n_2, k_2 \rangle|^2 $.
If there is only one band extremum($N^*=1$), (\ref{eq:2by2_formula_1}) always applies.
Since $d^{n_1, k_1}_{n_2, k_2} = d^{n_2, k_2}_{n_1, k_1}$ and $d^{n, k}_{n, k} =0$, we have at least $d^{\alpha, k_1}_{\alpha, k^*} \sim (k_1 - k^*)^2$ so that the sum in (\ref{eq:2by2_formula_1}) is converging for the quadratic band dispersion near $k^*$.
On the other hand, if $\mathrm{det}\mathcal{D}^{(1)}_{\alpha} \neq 0$, which is possible for $N^* \geq N_\mathcal{V}$, it becomes
\begin{align}
A_1(\varepsilon^\alpha) \approx \frac{1}{2\delta\varepsilon^2}\sum_{l_1}\sum_{l_2} d^{\alpha, k^*_{l_1}}_{\alpha, k^*_{l_2}}.\label{eq:2by2_formula_2}
\end{align}
From the sign of (\ref{eq:2by2_formula_1}) or (\ref{eq:2by2_formula_2}), one can calculate the Z$_{2}$ numbers $P$ and $P_\mathrm{half}$ for the $N_\mathcal{V} = 2$ case.

While we have assumed that the valence or conduction band is nondegenerate, the generalization to degenerate cases is straightforward: we need one more summation over the band index for the bands with the valence band maxima or conduction band minima in addition to the sum over $k^*_l$.
Details of the derivation of (\ref{eq:2by2_formula_1}) and (\ref{eq:2by2_formula_2}) are given in App.~\ref{app:N_V=2}.

\section{Examples for applications of \textit{\textbf{M}}, \textit{\textbf{P}}, and \textit{\textbf{P}}$_\mathrm{\mathbf{half}}$}\label{sec::examples}
To elucidate the above discussion, we directly apply these notions in the context of specific examples. Apart from the well-known standard symmetry protected topological models, the above general evaluations of the numbers $M$, $P$, and $P_\mathrm{half}$ also apply directly to graphene edges as well as grain boundaries.

\subsection{Rice-Mele model}\label{sec:rm}

In this example, we show that our bulk numbers predict the correct number of in-gap modes independent of the edge profile. This is in contrast to the Zak phase, which is the conventional topological invariant for 1D reflection symmetric insulators,  as its BBC is sensitive to whether the finite system is commensurate with the bulk unit cell or not\cite{Rhim2017}.

Consider in this regard the Rice-Mele model\cite{Rice1982}. This model entails  a 1D tight binding model consisting of two sites in the unit cell with a single orbital per site.
The left (right) site in the unit cell, denoted with A(B), has the onsite energy $\Delta$($-\Delta$).
The hopping between the nearest neighbor sites in the same (different) unit cell is $-t-\delta$($-t+\delta$).
Concretely,  the Hamiltonian is given by
\begin{align}
\mathcal{H}_{\mathrm{RM}}(k) = \bpm \Delta & s(k)e^{-i\phi_k} \\ s(k)e^{i\phi_k} & -\Delta \epm \label{eq:RM_ham}
\end{align}
where $s(k) =\sqrt{2(t^2+\delta^2) + 2(t^2-\delta^2)\cos k}$, and $e^{i\phi_k} = -2 (t\cos k/2 - i\delta\sin k/2)/s(k)$.
The energies are given by $\epsilon_{n,k} = (-1)^n[s(k)^2 + \Delta^2]^{1/2}$ where $n=1$ and $n=2$ represent the lower and upper bands, respectively.

\begin{figure}
	\begin{center}
		\includegraphics[width=\columnwidth]{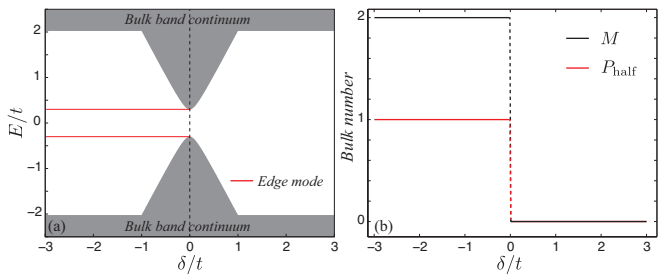}
	\end{center}
	\caption{(a) The band structure of the finite Rice-Mele chain with 1000 unit cells as a function of $\delta/t$. Here, $\Delta/t = 0.3$. Gray regions are bulk band continua. When $\delta/t < 0$, we have two edge modes at $E/t = \pm\Delta$ represented be red solid lines. (b) A plot of bulk numbers $M$ and $P_\mathrm{half}$ as a function of $\delta/t$. The obtained numbers are in direct correspondence, see Sec. \ref{sec:rm}.
	}
	\label{fig:RM}
\end{figure}

First, let us consider the number of edge modes when the terminated system is commensurate with the bulk unit cell, i.e., the total number of sites is even.
In this case, one can check that the number of edge states in the gap is two for $\delta/t < 0$, and zero for $\delta/t > 0$ as shown in Fig.~\ref{fig:RM}(a).
The edge states in the region where $\delta/t <0$ can disappear when they touch the bulk band's edge at $\delta/t =0$ although the bulk gap is never closed for finite $\Delta$.
Based on Sec.~\ref{sec:z_number}, one can readily determine the integer valued number $M$.   This confirms that the outlined procedure predicts the number of edge modes correctly as detailed in Fig.~\ref{fig:RM}(b) by the black lines.

On the other hand, the Z$_{2}$ number $P_\mathrm{half}$ is evaluated analytically as
\begin{align}
P_\mathrm{half} = \frac{1}{i\pi}\ln\mathrm{sgn}\left(t\delta\right).
\end{align}
Hence, if $t\delta < 0$ ($t\delta > 0$), we have an odd (even) number of edge modes in each of the upper and lower half of the gap. This is again consistent with the results for the finite-size system calculations.
The derivation of the above formula is detailed in App.~\ref{app:rice-mele}.

For the reflection symmetric case $\Delta=0$, where the Rice-Mele model reduces to the SSH model\cite{Su1979}, one can see that our Z$_{2}$ number reduces to the Zak phase\cite{Zak1989} in this commensurate case, which is the topological invariant for the 1D reflection symmetric insulators.
Since the wave function of the lower band reduces to $1/\sqrt{2}\bpm -1 & e^{i\phi_k} \epm^\mathrm{T}$, the parity of the wave function is $\xi_-(0) = t/|t|$ at $k=0$, and $\xi_-(\pi) = \delta/|\delta|$ at $k=\pi$, where $\xi_-(k)$ is the parity eigenvalue at $k$ of the lower band.
As a result, the Z$_{2}$ number is rewritten as
\begin{align}
P_\mathrm{half} = \frac{1}{i\pi}\ln\mathrm{sgn}\left(\xi_-(0)\xi_-(\pi)\right).
\end{align}
Since the Zak phase\cite{Zak1989} $\gamma = i\sum_{\mathrm{occ.}} \int_{\mathrm{BZ}} dk \langle u_{n,k} |\partial_k | u_{n,k} \rangle $, where $u_{n,k}$ is the cell-periodic part of the Bloch wave function and the summation is over the occupied bands, can be represented as 
$e^{i\gamma} = \prod_{\mathrm{occ.}}\xi_n(0)\xi_n(\pi)$ for reflection symmetric insulators\cite{Miert2017}, we arrive at 
\begin{align}
P_\mathrm{half} = \gamma/\pi.
\end{align}

As a next step, we consider incommensurate terminations where the finite chain consists of an odd number of sites. This ensures that there is no bulk unit cell commensurate with the terminated system.
The according termination is realized by applying an infinite onsite potential at one site given by
\begin{align}
\mathcal{V}_b = \lim_{V_0\rightarrow\infty}V_0 a_1^\dag a_1 \quad\mathrm{or}\quad \lim_{V_0\rightarrow\infty}V_0 b_1^\dag b_1
\end{align}
where $a_1^\dag$ and $b_1^\dag$ are the creation operators at the A and B site in the first unit cell.
This makes all orbitals at those sites irrelevant to the states within the bulk band's bandwidth by placing them at infinite energy, which effectively removes the A or B site.
Then, the pole-determinant is given by
\begin{align}
A_\beta(\varepsilon) = 1 - \beta V_0 \sum_{n,k} \frac{|\langle i | n,k\rangle |^2}{\varepsilon - \epsilon_{n,k}} \label{eq:RM_A_odd}
\end{align}
where $|i\rangle$ is $a_i^\dag|0\rangle$ or $b_i^\dag|0\rangle$.
Note that the coefficient of $\beta$ in (\ref{eq:RM_A_odd}) is positively (negatively) divergent near the conduction (valence) band since $|\langle i | n,k\rangle |^2$ is positive.
Therefore, one cannot have zeros for $A_\beta(\varepsilon^c)$ at the conduction band portal, and $M$ in (\ref{eq:Z_number}) has contributions only from the valence band ($\alpha=v$). 
Since $\partial_\varepsilon A_\beta(\varepsilon) = \beta V_0 \sum |\langle i | n,k\rangle |^2/(\varepsilon - \epsilon_{n,k})^2$ is positive, we have $M = P =  1$ for any values of tight binding parameters.
This is precisely consistent with the finite-size system calculations where we always find a single edge mode in the gap for finite gap. 
We emphasize that this cannot be predicted from the conventional bulk number, i.e., the Zak phase.
While one has different Zak phases depending on the relative sign between $t$ and $\delta$, one can find the valid BBC only for the commensurate cases as in the first case in the above.
In addition, the Zak phase's BBC can only be applied to reflection-symmetric cases ($\Delta = 0$).
On the other hand, $M$ and $P$ can be applied to arbitrary insulators without any of the symmetry restrictions or the commensurability issues.

\begin{figure}
	\begin{center}
		\includegraphics[width=\columnwidth]{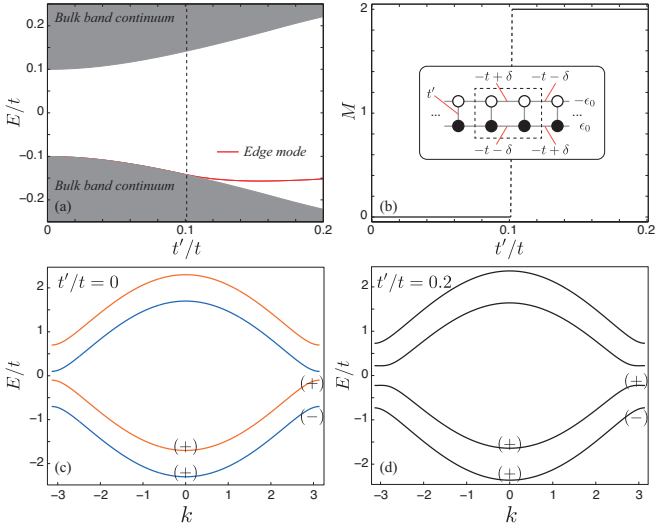}
	\end{center}
	\caption{(a) The band structure of the finite double SSH chain with 1000 unit cells as a function of $t^\prime/t$ where $t^\prime$ is the inter-chain coupling. Here, $\delta/t = 0.2$, and $\epsilon_0/t = 0.3$. Gray regions are bulk band continua, and red solid curves  are doubly degenerate edge states. (b) The bulk number $M$ for the central gap as a function of $t^\prime/t$. The inset describes the double SSH chain where the dashed box is the unit cell. In (c) and (d), we plot the band structures for  $t^\prime/t = 0$ and $t^\prime/t =0.2$. When $t^\prime/t = 0$, the upper and lower SSH chains in the inset of (b) are decoupled and their bands are drawn by blue and yellow curves in (c). Finally,  the parity of the wave function at $k=0$ and $k=\pi$ by $(\pm)$ is marked.
	}
	
	\label{fig:dssh}
\end{figure}

\subsection{Double Su-Schrieffer-Heeger model}

We may accordingly examine another 1D example, where the Zak phase's BBC fails even though the finite system is {\it commensurate} with the bulk unit cell.
This is because the system becomes metallic as a result of the termination\cite{Rhim2017,Vanderbilt1993b}.
Nonetheless, the above defined bulk numbers once again are fully compatible.

Specifically, consider two coupled SSH chains as illustrated in the inset of Fig.~\ref{fig:dssh}(b).
The hopping parameter is $-t-\delta$ between neighboring sites in the same unit cell, and $-t+\delta$ between those in the different unit cell for the lower chain, and vice versa for the upper chain.
The inter-chain coupling is represented by $t^\prime$, and those two chains have different onsite potentials, $\pm\epsilon_0$.

One can immediately note that, when $t^\prime = 0$,  there are no edge modes for sufficiently small onsite potentials even though the Zak phase is nontrivial.
In this case, the lower SSH chain is trivial ($\gamma =0$) while the upper one is nontrivial ($\gamma =\pi$) when $\delta/t$ is positive as shown in the previous section.
The full band structure for this case is plotted in Fig.~\ref{fig:dssh}(c) where red (blue) curves come from the trivial (nontrivial) chain.
For commensurate finite systems, two edge modes will be generated between two nontrivial bands from the upper chain.
However, those edge modes fall into the lower bulk band of the lower chain if the onsite energy difference is not large enough.
As a result, one does not have in-gap modes in the central gap of the whole double chain system as plotted in Fig.~\ref{fig:dssh}(a) although the Zak phase is nontrivial as clear from the parity configurations at reflection symmetric momenta in Fig.~\ref{fig:dssh}(c) and (d).
This situation remains up to a critical inter-chain coupling and, above it, the conventional Zak phase's BBC starts to hold.
The reason for this mismatch lies in fact that  the system turns metallic after the termination, while both the bulk and the finite system should be insulating for the application of the Zak phase's BBC.

The Z$_{2}$ number $P$ for the band gap between the second and third bands can readily be determined to be even.
Although this is consistent with the finite size system calculations, we cannot completely determine the existence of the edge modes with only this number.
Since the double SSH model does not have chiral symmetry, as manifested by the edge mode's spectrum in Fig.~\ref{fig:dssh}(a), one cannot apply the chiral Z$_{2}$ number $P_\mathrm{half}$.
However, our number $M$ exactly predicts the number of edge modes in the central gap precisely as presented in Fig.~\ref{fig:dssh}(b).

\begin{figure}
	\begin{center}
		\includegraphics[width=\columnwidth]{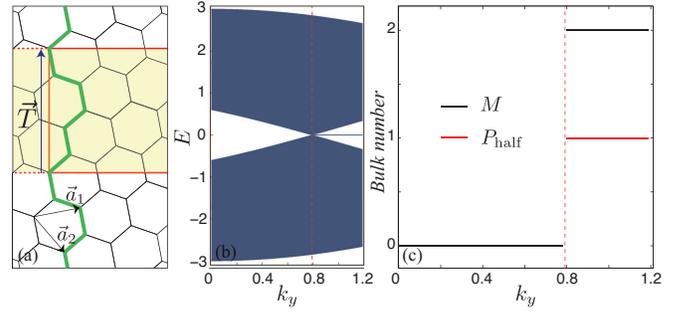}
	\end{center}
	\caption{(a) The lattice structure of graphene with lattice vectors $\vec{a}_1$ and $\vec{a}_2$. We consider the GNR with the lattice vector $\vec{T}$ along the $y$ -axis. The boundary is represented by the thick green line. In the outlined procedure, one considers the enlarged unit cell ( the red solid box), containing 28 sites. For a given $k_y$, we 
have an effective 1D system along the $x$- axis as represented by the infinite yellow region. (b) Plot of the band structure of the GNR with 1680 sites in the GNR's unit cell. The red dashed line represents the position of the Dirac point at $k_y = 2\pi/3\sqrt{7}$. (c) The two bulk numbers $M$ and $P_\mathrm{half}$ plotted as a function of $k_y$. Again direct correspondence is retrieved.
}
	\label{fig:graphene}
\end{figure}

\subsection{Graphene nanoribbons}

In this section, we show that our bulk number $M$ predicts the exact number of edge modes of graphene nanoribbons (GNRs) as a function of the preserved momentum. 
Previously, Delplace et~al.\ demonstrated that the Zak phase $\gamma$ can serve as  a good bulk number, with $\gamma =0$ and $\gamma =\pi$ corresponding to the nonexistence and existence of edge modes respectively\cite{Delplace2011}.
However, this evaluation has only be made explicit for the zigzag GNR.
While it works correctly due to the Z$_{2}$ nature of the Zak phase, one can however not confirm the nonexistence of the edge modes when $\gamma =0$ a priori.
Similarly, this correspondence cannot be applied to the mixed edges such as the GNR with zigzag edge on the left and the bearded edge on the right edges.
In contrast, the bulk number $M$ predicts precisely the number of edge modes of GNRs with {\it arbitrary} cutting direction and edge profiles. 
These findings are particularly interesting from an experimental point of view because the existence of edge states usually gives rise to the magnetic order which might lead to the spintronics applications.
Indeed, due to the ready availability of graphene, there exist experimental studies on the edge states of graphene\cite{Cai2010,Tao2011,Wang2016}.

Recall that in graphene the three nearest neighbor vectors from A to B sites are given by $\boldsymbol\delta_1 = a/\sqrt{3}( 1/2,\sqrt{3}/2 )$, $\boldsymbol\delta_2 = a/\sqrt{3} (-1,0) $, and $\boldsymbol\delta_3 = a/\sqrt{3}( 1/2,-\sqrt{3}/2 )$, where $a$ is the lattice constant.
The Hamiltonian reads
\begin{align}
\mathcal{H}_\mathrm{graphene} = \bpm 0 & f(k_x,k_y) \\ f(k_x,k_y)^* & 0 \epm
\end{align}
where $f(k_x,k_y) = \sum_i e^{i\v k\cdot\boldsymbol\delta_i} = -2\cos(a k_y/2) e^{ia k_x/2\sqrt{3}} - e^{-ia k_x/\sqrt{3}}$.

\begin{figure}
	\begin{center}
		\includegraphics[width=\columnwidth]{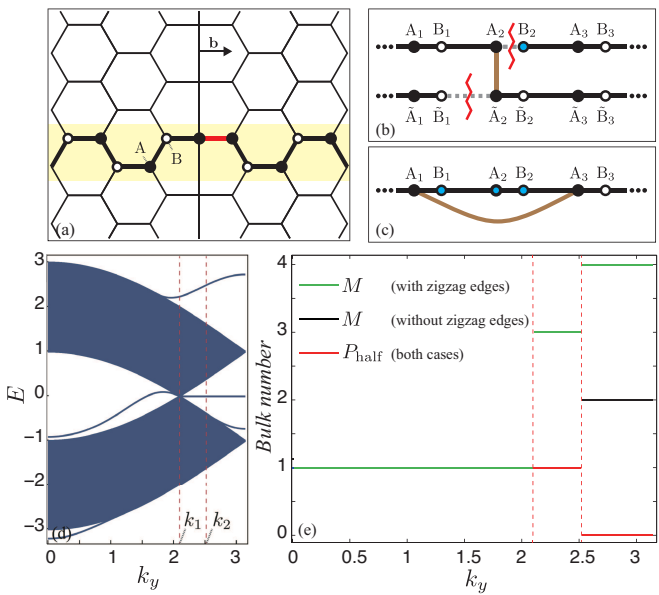}
	\end{center}
	\caption{(a) Structure of the ``55" grain boundary of graphene characterized by the Burgers vector $\mathbf{b}$. The effective 1D model for given $k_y$ is represented be the yellow stripe. The two sublattices are labeled by A (filled circle) and B (open circle). In (b) and (c) we present two different ways of making the effective ``55" grain boundary. In (b), one first imagines preparing two 1D effective chains (upper and lower ones) for graphene. Then,  the bonds along A$_2$B$_2$ as well as $\tilde{\mathrm{B}}_1$$\tilde{\mathrm{A}}_2$ are disconnected  and the B$_2$ site is removed  by applying a large onsite potential. Finally, the upper and lower chains are reconnected by the red vertical bond between A$_2$ and $\tilde{\mathrm{A}}_2$, which correspond to the red bonds in (a). In this case, one retrieves zigzag edges at the outer boundaries. On the other hand, as described in (c), one can also make the ``55" grain boundary without additional zigzag edges. To this end, one starts from a single 1D effective chain. Then, three sites between A$_1$ and A$_3$ are removed by applying a large onsite potential. Finally, one creates a bond between A$_1$ and A$_3$ as indicated by the red solid curve. Since the left- and right-ends of this 1D effective chain are connected in the ring geometry, the systems does not have any edges other than the ``55" grain boundary in the middle. (d) The band structure of the GNR with the grain boundary in the middle and two zigzag edges at the ends. $k_1$ is the Dirac point and $k_2$ entails the momentum at which one bulk state starts coming into the central gap. (e) The four describing bulk numbers are plotted as a function of $k_y$. The green and black ones comprise the $\mathbf{Z}_\geq$ number for terminations (b) and (c). 
	The  $\mathbf{Z}_2$ number is the same for both cases. As before the retrieved bulk numbers are in direct correspondence.
	}
	\label{fig:55grain}
\end{figure}

Let us first consider the zigzag GNR where the atoms at the left edge belong to the A sublattice and those at the right edge are belong to B sublattice.
For given $k_y$, $f(k_x,k_y)$ can be interpreted as a Hamiltonian matrix's element of a fictitious 1D chain which contains two sites in the unit cell.
The phase factors $e^{-ia k_x/2\sqrt{3}}$ and $e^{-ia k_x/\sqrt{3}}$ can be interpreted as the Bloch phase difference for the intracell and the intercell hoppings of this 1D chain model along the $x$ axis with corresponding hopping parameters $-2\cos ak_y/2$ and $-1$.
If we consider a finite version of this 1D model, it is identical to the commensurately terminated  SSH model with $-t-\delta = -2\cos ak_y/2$ and $-t+\delta = -1$.
Thus, we can apply the bulk numbers of the SSH model to this effective 1D model of the zigzag GNR. This leads to the following criterion for the existence of edge modes for a given $k_y$.:
\be
\left( 2\cos\frac{ak_y}{2} \right)^2 -1 < 0.
\ee
Accordingly, we conclude that there are two edge modes when $-\pi < ak_y < -2\pi/3$ and $2\pi/3 < ak_y < \pi$ for the zigzag GNR as is consistent with previous studies\cite{Fujita1996,Ryu2002}.

As another kind of the termination, we consider the bearded GNR along the $y$ direction.
In this case, the position of the above A and B sites are reversed, so that the intracell and intercell hoppings now become $-t+\delta = -2\cos ak_y/2 $ and $-t-\delta = -1$.
As a result, we have two edge states when $-2\pi/3 < ak_y < 2\pi/3$ which is again consistent with previous work\cite{Ryu2002}.

We can also deal with a GNR having a zigzag edge on one side and a bearded edge on the other side.
In this case, the effective finite 1D system for given $k_y$ is equivalent to the incommensurately terminated SSH chain in Sec.~\ref{sec:rm}. Hence there exists a single edge mode in the gap regardless of the tight binding parameters as long as the effective 1D system for given $k_y$ is insulating.

For the armchair GNR, the effective 1D Hamiltonian for given $k_x$ consists of four basis sites in the unit cell. Using standard numerical means, 
we find all the bulk numbers $M$, $P$ and $P_\mathrm{half}$ are zero for all momenta which is consistent with the absence of edge modes in the armchair GNR\cite{Fujita1996,Ryu2002}, see Appendix~\ref{app:agnr} for details.

Finally, our bulk numbers $M$ and $P_\mathrm{half}$ can be calculated for arbitrary directions with arbitrary edge profiles.
As an example, we consider a GNR with an edge shape shown in Fig.~\ref{fig:graphene}(a). That is,  the lattice vector is $\vec{T} = 2\vec{a}_1 - 3\vec{a}_2$.
To study this GNR, we assume the enlarged unit cell for graphene as represented by the red solid box in Fig.~\ref{fig:graphene}(a) which contains 28 sites.
For given momentum $k$ along $\hat{y}$, we have an effective 1D system with this unit cell.
The yellow region is a part of this 1D system in which the dashed and the solid boxes are the $(m-1)$-th and the $m$-th unit cells of it.
In this effective 1D model, we calculate $M$ and $P_\mathrm{half}$ to predict the number of edge modes in the central gap when the GNR is terminated as in Fig.~\ref{fig:graphene}(a) by the thick green line.
We obtain $M = 2$, $P_\mathrm{half}=1$ when $-\pi/\sqrt{7} < ak < -2\pi/3\sqrt{7}$ and $2\pi/3\sqrt{7} < ak < \pi\sqrt{7}$, and $M = P_\mathrm{half}=0$ otherwise. This is once more consistent with the band structure of this GNR (see Fig.~\ref{fig:graphene}(b)). 

\subsection{Graphene's ``55" grain boundary }\label{sec:grain_boundary}

One can also apply our bulk numbers to predict the number of localized modes around a junction between two systems by choosing an appropriate local operator $\mathcal{V}_b$. 
This is, not in the least place, directly relevant in the context of experiments. 
Indeed, quite some recent works have found surface states at graphene defects, including graphene grain boundaries\cite{Lahiri2010,Kim2011,Huang2011,Tsen2012,Biro2013,Phillips2015}.
As an example, let us consider the ``55" grain boundary in graphene which is characterized by the Burgers vector $\mathbf{b} = a/\sqrt{3}\hat{x}$ as shown in Fig.~\ref{fig:55grain}(a)\cite{Phillips2015}.
Since we cut the system in the zigzag direction, we thus first obtain the same effective 1D effective chain of the zigzag GNR detailed in the previous subsection.
One can then realize the grain boundary (GB) in two ways as follows.

The first one corresponds to Fig.~\ref{fig:55grain}(b). 
The system is in this case prepared using two identical SSH-like models on two ring geometries.
The bonds between the A sites (the red bond in Fig.~\ref{fig:55grain}(a)) are obtained  by eliminating connections between A$_2$, B$_2$ in one chain and $\tilde{\mathrm{B}}_1$,$\tilde{\mathrm{A}}_2$ in  the other chain. The two dangling sites A$_2$ and $\tilde{\mathrm{A}}_2$ are then reconnected.
While the B$_2$ and $\tilde{\mathrm{B}}_1$ sites correspond to opposite edges, we remove the B$_2$ site by putting a large potential $V_0$ on it.
The local operator is thus given by
\begin{align}
\mathcal{V}_b =& \left( 2\cos\frac{k_y}{2}c^\dag_{\mathrm{A}2}c_{\mathrm{B}2} + c^\dag_{\tilde{\mathrm{B}}1}c_{\tilde{\mathrm{A}}2} -c^\dag_{\mathrm{A}2}c_{\tilde{\mathrm{A}}2} + \mathrm{h.c.} \right) \nonumber \\
& + \lim_{V_0 \rightarrow \infty} V_0c^\dag_{\mathrm{B}2}c_{\mathrm{B}2}.
\end{align}
This junction represents the finite GNR with the ``55" grain boundary in the middle, and zigzag edges at outer boundaries.

The second manner of creating the  ``55" GB departs from a single SSH chain as shown in Fig.~\ref{fig:55grain}(c).
In this case, the red bonds between the A sites are realized by getting rid of the B$_1$, A$_2$, and B$_2$ sites. This is done  by means of putting a large onsite potentials on them. As a next step, the two dangling sites A$_1$ and A$_3$     are then reconnected. 
The local operator in this case reads
\begin{align}
\mathcal{V}_b =& \lim_{V_0 \rightarrow \infty} V_0 \left( c^\dag_{\mathrm{B}1}c_{\mathrm{B}1} + c^\dag_{\mathrm{A}2}c_{\mathrm{A}2} + c^\dag_{\mathrm{B}2}c_{\mathrm{B}2} \right) \nonumber \\
& - \left( c^\dag_{\mathrm{A}1}c_{\mathrm{A}3} + c^\dag_{\mathrm{A}3}c_{\mathrm{A}1} \right).
\end{align}
Since we are dealing with a ring geometry, the grain boundary made in this way is the only boundary, and there no outer boundaries.

The band structure of the finite width GNR with such a grain boundary  is plotted in Fig.~\ref{fig:55grain}(d).
Firstly, we note that we have a different number of in-gap surface bands depending on $k_y$;
(i) for $0\leq k_y \leq k_1$ there is only one surface band, whose states are localized at the grain boundary, (ii) for $k_1 \leq k_y \leq k_2$ there are three zero-energy flat surface bands. The according wave functions of two of them are localized at outer zigzag edges, whereas the wave function of the remaining one is localized at the grain boundary, (iii) for $k_2 \leq k_y \leq \pi$ we retrieve four surface bands. The profiles of the three flat ones are the same as those of (ii), and the remnant dispersive one has wave functions localized around the grain boundary.
These features are directly consistent with our bulk numbers as detailed in Fig.~\ref{fig:55grain}(e).
With the configuration in Fig.~\ref{fig:55grain}(b), the bulk number $M$ counts the total number of in-gap modes correctly as drawn by the green lines in Fig.~\ref{fig:55grain}(e).
On the other hand, $M$ of another termination (Fig.~\ref{fig:55grain}(c)) predicts only the number of surface modes localized at the grain boundary accurately as shown by the black lines in Fig.~\ref{fig:55grain}(e).
The $\mathbf{Z}_2$ number $P_\mathrm{half}$, which is the same for both terminations, finally also yields the correct even-odd prediction of the number of in-gap modes as presented by the red curves in Fig.~\ref{fig:55grain}(e).

\begin{figure}
	\begin{center}
		\includegraphics[width=0.75\columnwidth]{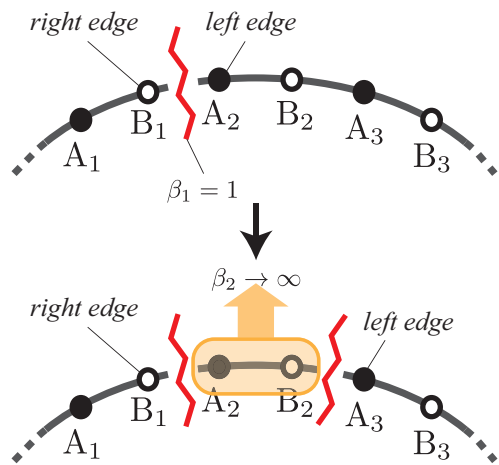}
	\end{center}
	\caption{A cartoon for the effective pole winding periodic process. We apply local potentials $\beta_1\mathcal{V}_1$ and $\beta_2\mathcal{V}_2$ in turn where $\mathcal{V}_1$ makes an open boundary between the sites B$_1$ and A$_2$, and $\mathcal{V}_2$ is and onsite potential on the sites A$_2$ and B$_2$. Starting from $\beta_1 =1$ and $\beta_2=0$ (the open boundary between B$_1$ and A$_2$), the same system at $\beta_2 = \infty$ is obtained in the thermodynamic limit except for the irrelevant states in the yellow box that have infinite energy as $\beta_2 = \infty$. One can deal in this setup with general cases with arbitrary number of orbitals and basis sites in the same way. 
	 }
	\label{fig:eff_periodic}
\end{figure}

\section{Pole winding number and chirality for higher dimensions}\label{sec::pole}

We can also apply the numbers for 1D insulators in the previous sections to higher dimensions by defining an effective 1D Hamiltonian obtained by performing the Fourier transformation only along the directions parallel to the edge or surface to be made like the graphene examples in the previous section.
The effective 1D Hamiltonian is characterized by the momentum parallel to the edge or surface, and we can investigate the number of in-gap modes by scanning the whole parallel momenta.
However, in higher dimensions, it is desired to know the existence of the chirality or helicity of boundary modes which traverse the gap from the valence band to the conduction band without disconnections.
To this end, in this section, we define a winding number, which we call the \textit{pole winding number}, for 2D from bulk wave functions by analyzing the behavior of  poles of the Green's function near the band edges.
Then, we discuss how to apply this number to 3D insulators.

\begin{figure}
	\begin{center}
		\includegraphics[width=0.75\columnwidth]{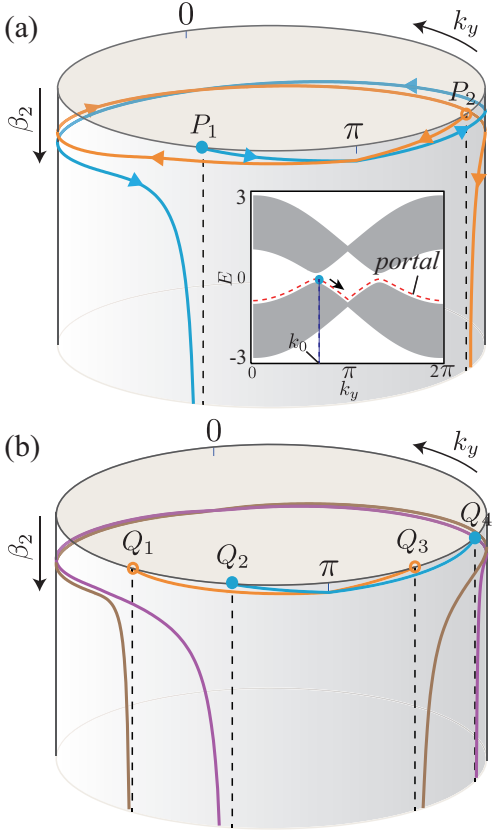}
	\end{center}
	\caption{(a) Poles' evolutions for the topologically nontrivial phase of the Kane-Mele model in the $k_y\beta_2$-space portrayed as cylindrical surface. Here we set $\lambda_v = 0.1t$, $\lambda_\mathrm{SO} = -0.06t$, and $\lambda_\mathrm{R} = 0.05t$. We assume that $\hat{z}$ is downward. $k_y$ is the momentum parallel to the edge of the 2D system. In the inset, we denote the valence band portal line by the red dashed curve, and the starting pole at $k_0$ when $\beta_2 = 0$. (b) Pole's evolutions for the topologically nontrivial phase when $\lambda_v = 0.4t$. In both (a) and (b), $P_i$'s and $Q_i$'s represent starting poles.
	}
	\label{fig:winding_km}
\end{figure}

\subsection{Effective periodic process}
To define this winding number, let us first introduce an \textit{effective} periodic process, which is depicted in Fig.~\ref{fig:eff_periodic}.
Dealing with 2D insulators with an arbitrary number of orbitals per site, this chain geometry in Fig.~\ref{fig:eff_periodic} represents its effective 1D system obtained by fixing a momentum parallel to the edge we are interested in.
We denote this momentum as $k_y$ without loss of generality.
Next, we consider two kinds of local operators $\beta_1\mathcal{V}_1$ and $\beta_2\mathcal{V}_2$, where $0\leq \beta_1 \leq 1$, and $0 \leq \beta_2 <\infty$.
Physically, $\mathcal{V}_1$ removes all the hopping processes across the boundary (red line) between B$_1$ and A$_2$, and $\mathcal{V}_2$ is the onsite potential for all the orbitals in the unit cell on the right-hand side of the boundary as shown by the yellow box in Fig.~\ref{fig:eff_periodic}.
Here, for concreteness, we specified $\mathcal{V}_2$ to be operating on the right-hand side unit cell to analyze edge modes localized to the left edge of the terminated system.
Starting from the bulk Hamiltonian $\mathcal{H}_0$, first, we make open boundaries at B$_1$(right edge) and A$_2$(left edge) by setting $\beta_1 = 1$.
Then, an effective periodic process is obtained by controlling $\beta_2$ from $0$ to $\infty$.
This simply moves the left edge from A$_2$ to A$_3$ which in the thermodynamic limit is exactly the same system obtained by the first operation($\beta_1 = 1$) except the existence of the orbitals in the sites between two new boundaries. 
However, those orbitals are at infinite potential and irrelevant to the in-gap surface modes we are focusing on, and this is why we call the second process an effective periodic process. 

\subsection{Pole winding number and chirality}
Let us set a portal in the gap around the valence band similar to the 1D case in the previous sections.
This becomes a line in the 2D case as illustrated by the red dashed line in the inset of Fig.~\ref{fig:winding_km}(a), and we call it the \textit{valence} or \textit{conduction band portal line}.
We can now show that the topology of the in-gap surface modes localized at the left edge is encoded in the evolution of the Green's function's pole along the valence band portal line during the effective periodic process.
This evolution can be represented by a parameterized curve in the 2D space of $k_y$ and $\beta_2$, the \textit{pole-curve} plotted by solid curves in Fig.~\ref{fig:winding_km}.
The pole-curve is obtained from
\begin{align}
0 = \mathrm{det}\left\{1 - \mathcal{G}_0(\varepsilon^v_{k_y})\mathcal{V}_1 - \beta_2\mathcal{G}_0(\varepsilon^v_{k_y})\mathcal{V}_2\right\}
\end{align}
where $\varepsilon^v_{k_y} = \epsilon_{k_y}^v + \delta\varepsilon$ is the valence band portal line with $\epsilon^v_{k_y}$ the valence band's edge and $\delta\varepsilon$ is an infinitesimal positive value.
Since $k_y=0$ and $k_y=2\pi$ are identified in the Brillouin zone (BZ), one can represent this 2D space as a surface of a semi-infinite cylinder with unit radius as illustrates in Fig.~\ref{fig:winding_km}.
While the pole-curves exhibited in Fig.~\ref{fig:winding_km} are obtained from a specific model, the Kane-Mele model, we discuss the general classification of the pole-curves referring to this model.

\begin{figure*}
	\begin{center}
		\includegraphics[width=2\columnwidth]{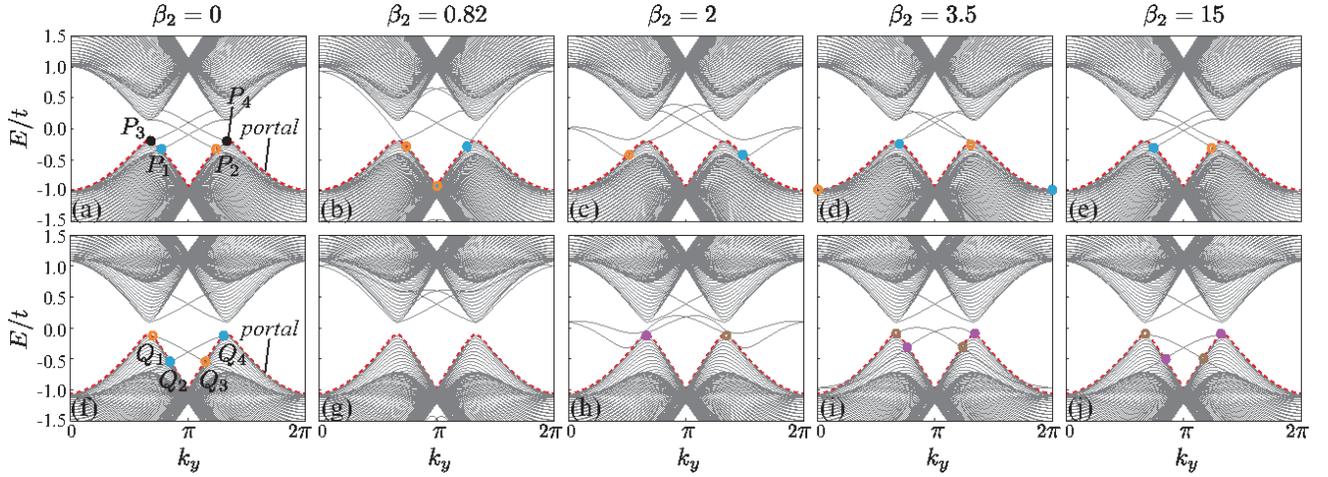}
	\end{center}
	\caption{Band spectra of the Kane-Mele model for varying $\beta_2$. The upper panels comprise topologically nontrivial cases having parameters $\lambda_v = 0.1t$, $\lambda_\mathrm{SO} = -0.06t$ and $\lambda_\mathrm{R} = 0.05t$. The lower panels depict trivial cases with $\lambda_v = 0.4t$, $\lambda_\mathrm{SO} = -0.06t$ and $\lambda_\mathrm{R} = 0.05t$. The valence band portals are represented by red dashed curves right above the valence band edges. In (a) and (f), the $P_i$'s and $Q_i$'s indicate the starting poles, which are denoted by the same markers as in Fig.~\ref{fig:winding_km}(b) and (c). From (b) to (e), the poles connecting  $P_i$'s are similarly marked. On the other hand, poles connected from the $Q_i$ points disappear in to the valence band continuum at $\beta_2 = 0.57$, and we have no poles at the valence band portal at $\beta_2 = 0.82$ as shown in (g). Above $\beta_2 = 2$, new poles are generated by the surface bands pulled out from the conduction band continuum as illustrated from (h) to (j). In both cases, the band structures at large $\beta_2$ return to original ones at $\beta_2=0$. The pole windings unambiguously discern the topological nature of the band structure.}
	\label{fig:pole_evolution}
\end{figure*}

While the pole's momenta at $\beta_2=0$ and $\beta_2 = \infty$ should be equal due to the effective periodic process, one can classify pole-curves into trivial and nontrivial cases as follows.
First, we denote the poles on the portal at $\beta_2 =0$ as the starting pole such as the $P_i$'s and $Q_i$'s in Fig.~\ref{fig:winding_km}.
If we represent one of the momenta of starting poles by $k_0$, the pole-curve starting from it is nontrivial if it connects two points $(k_y,\beta_2) = (k_0,0)$ and $(k_0,\infty)$ by winding the cylinder one or more times, being differentiable. 
The two pole-curves starting from $P_1$ and $P_2$, colored by blue and yellow in Fig.~\ref{fig:winding_km}(a), are nontrivial since they wind the cylinder once and it is differentiable through the whole curve.
This curve can be described by $\mathbf{p} = \hat{\rho} + \varphi\hat{\varphi} + \beta_2(\varphi)\hat{z}$ with the polar angle $\varphi$ as a parameter.
$\varphi$ runs from $k_0$ to $k_0 + 2\pi n_p$ during the effective periodic process where we call the integer $n_p$ the pole winding number.
Then, we define the chirality of the pole-curve as 
\begin{align}
\Xi = \frac{n_p}{|n_p|}
\end{align}
for $n_p\neq 0$, and $\Xi=0$ for $n_p = 0$.
If the pole winding number is nonzero and the chirality is positive(negative), we have one surface band connecting the valence and conduction bands with positive(negative) average velocity.
Note that the discussions so far correspond to only one edge as we consider one choice of the effective periodic process such as the one described in Fig.~\ref{fig:eff_periodic}.
If we consider another effective periodic process where the local potential $\mathcal{V}_2$ is applied to A$_1$ and B$_1$ sites, the pole winding number and chirality calculated from it characterize the in-gap modes localized to the opposite edge.

In contrast, the pole-curve is trivial in the following cases:
(i) If the pole-curve connects two poles at $\beta_2 = 0$ such as $Q_1$ and $Q_3$ or $Q_2$ and $Q_3$ in Fig.~\ref{fig:winding_km}(b), both ends of the corresponding surface band are connected to the valence band continuum and not considered robust.
(ii) If the pole-curve is a straight line that does not wind the cylinder; this corresponds to the edge state localized to the opposite edge of what we are interested in, and we consider it trivial with respect to the interested edge.
This is because both edges are separated from each other for $\beta_1=1$, and the local operator $\mathcal{V}_2$ cannot affect surface modes localized at this side.
As a result, the corresponding poles do not respond to the increase of $\beta_2$ and the pole-curve from these poles is just a straight line along $\hat{\beta_2}$ and cannot wind the cylinder.

Usually we have several poles on the portal line at $\beta_1=1$ and $\beta_2 = 0$, and pole-curves starting from them.
One can readily evaluate the pole winding number and the chirality for each of them.
From those topological numbers, one can analyze the structures of the in-gap modes.
If the sum of the chiralities is nonzero, the gap hosts chiral edge states as in the case of the quantum anomalous Hall effect.
Even if the total chirality vanishes, if there is a pole-curve with nonzero chirality, the system exhibits helical edge states like in the quantum spin Hall effect.
To further determine whether the winding number and chirality is robust against a certain symmetry breaking (e.g., time-reversal symmetry for the quantum spin Hall effect), we should evaluate different pole winding numbers and chiralities with the replacement of $\mathcal{V}_2$ by another potential that breaks the symmetry.
This new local potential would open the symmetry protected band crossings which is reflected in disconnected pole-curves.

While the above analysis on the topological characters makes the application of the formalism insightful, one can gain a more concrete understanding by observing evolutions of in-gap surface bands of the model with zigzag open boundaries as follows.
First, the pole winding for the nontrivial case is described from Fig.~\ref{fig:pole_evolution}(a) to (e).
At $\beta_2 = 0$, which is the starting point of the effective periodic process, there are four starting poles $P_i$'s.
Wave functions in surface bands starting from $P_1$ and $P_2$ are localized on the left edge while those from $P_3$ and $P_4$ are localized on the right edge.
Since $\mathcal{V}_2$ operates only on the left edge, and the two opposite edges are detached for $\beta_1=1$, surface bands connected to $P_3$ and $P_4$ do not respond to the increase of $\beta_2$ as shown in Figs.~\ref{fig:pole_evolution}(a) to (e).
As a result, pole-curves starting from them never wind the cylinder in Fig.~\ref{fig:winding_km}.
Since we are investigating pole winding numbers for the left edge, we can neglect them and only consider pole-curves starting from $P_1$ and $P_2$.
Surface bands connected to $P_1$ and $P_2$ move down as $\beta_2$ grows, and the poles for those bands also traverse along the valence band portal.
If we plot those poles in the $k_y \beta_2$ plane, as portrayed on the cylindrical surface in Fig.~\ref{fig:winding_km}(b), they travel the whole BZ once in opposite direction.
In other words, the pole winding numbers for the pole-curves starting from $P_1$ and $P_2$ are 1 and $-1$.
This is because bands cannot be disconnected abruptly at a certain momentum, and the poles should come back to their original position at $\beta_2=\infty$.

On the other hand, for the trivial phase, described in lower panels of Fig.~\ref{fig:pole_evolution}, there are no surface bands connecting the valence and conduction bands.
As a result, any of the starting poles $Q_i$'s, which are all left edge localized, cannot be continuously connected from the poles at $\beta_2 >2$.
To return to the band structure started with, new surface bands come down from the conduction band as shown in Fig.~\ref{fig:pole_evolution}(h), and it generates new pole curves independent of those in Fig.~\ref{fig:pole_evolution}(f).
Those poles from the new surface bands, marked by purple and brown circles from Fig.~\ref{fig:pole_evolution}(h) to (j), make pole-curves with the same colors in Fig.~\ref{fig:winding_km}(b).
Consequently, no poles starting from $Q_i$'s can wind the BZ completely which means pole winding numbers are all zero.

\begin{figure}
	\begin{center}
		\includegraphics[width=\columnwidth]{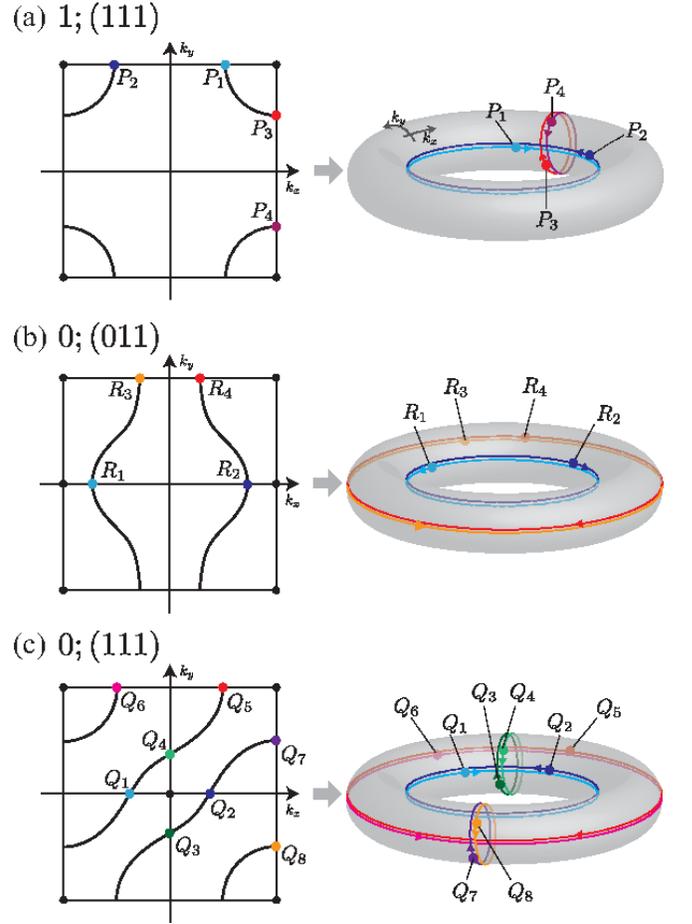}
	\end{center}
	\caption{Pole windings on the 2D torus BZ for various TRS TIs with $\nu_0;(\nu_1,\nu_2,\nu_3)=1;(111)$, $0;(011)$, and $0;(111)$. In the left panels, black solid curves are the starting pole lines projected onto the 2D BZ, and $P_i$, $R_i$, and $Q_i$ are the starting poles that yield nontrivial winding numbers. Dirac points are marked by black dots. In the right panels, we show how the poles starting from those starting poles wind the BZ torus during the effective periodic process.}
	\label{fig:winding_3d}
\end{figure}

\subsection{3D applications of the pole winding number}
The pole winding number can also be applied to 3D cases.
Let us consider the BBC for the surface perpendicular to $\hat{z}$ so that the surface BZ is the $k_x$$k_y$-plane. 
In 3D, the valence band portal should be a 2D curved surface described by $E = S(k_x,k_y)$ which is the surface slightly above the valence band edge.
If we fix $k_x$($k_y$), we have an effective valence band portal line on a 1D sub-BZ along $k_y$($k_x$). 
Along this portal line we can evaluate pole winding numbers as in the previous subsection.
When $\beta_1 =1$ and $\beta_2 = 0$, the starting poles on the valence band portal usually consist of lines because they are just the intersection of two surfaces, $E = S(k_x,k_y)$ and the surface bands in the gap.
We investigate the surface band structure by considering the pole winding numbers for the distinct cases of how the effective valence band portal lines touch the starting pole lines in the surface BZ
In the following, we show how this could be done for 3D TIs as an example.

First, for the strong  or TRS protected TI, we usually have a Dirac cone as a surface band in the gap.
In this case, the starting poles form a circle enclosing the Dirac cone as presented in Fig.~\ref{fig:winding_3d}(a). 
In Fig.~\ref{fig:winding_3d}, the starting pole lines, the black solid curves, are projected onto the $k_x k_y$-plane, and the black dots at the high symmetry points are the Dirac points.
For the case of the strong TI, only two effective valence band portal lines, $S(\pi,k_y)$ and $S(k_x,\pi)$ offer nontrivial pole winding numbers because the protected surface bands' crossings along them (at $\mathbf{k}=(\pi,\pi)$) would be manifested as this number.
On the torus manifold of the 2D BZ in Fig.~\ref{fig:winding_3d}(a), the poles starting from $P_1$($P_3$) and $P_2$($P_4$) wind the surface of the torus following the non-contractible loops along the toroidal (poloidal) direction with opposite chiralities to each other.
This is the unique feature of the Dirac surface band from the pole winding, and if one observe the odd number of starting poles with the positive chirality and the same number of ones with the negative chirality for both toroidal and poloidal directions in all surfaces of the TRS insulator, this phase is the strong TI.

On the other hand, for a weak TI\cite{note1} which has even number of Dirac cones in the surface BZ, the starting pole lines appear to be traversing the 1D sub-BZ as shown by the black curves in Fig.~\ref{fig:winding_3d}(b) and (c). We can then define weak invariants $\nu_0;(\nu_1,\nu_2,\nu_3)$.
For $\nu_0;(\nu_1,\nu_2,\nu_3) = 0;(011)$, there are two effective valence band portal lines, $S(k_x,0)$ and $S(k_x,\pi)$, that carry the nontrivial pole winding along the toroidal direction, and we have two starting poles $R_1$ and $R_3$ with the negative chirality, and the other two $R_2$ and $R_4$ with the positive chirality as shown in Fig.~\ref{fig:winding_3d}(b).
For another kind of weak TI with $\nu_0;(\nu_1,\nu_2,\nu_3) = 0;(111)$, one can find nontrivial pole windings along four effective valence band portal lines $S(k_x,0)$, $S(k_x,\pi)$, $S(0,k_y)$, and $S(\pi,k_y)$.
We have positive chirality for the pole windings starting from $Q_{2n-1}$'s, and negative for $Q_{2n}$'s where $n$ is an integer from 1 to 4.
In conclusion, if the TRS insulator has even number of starting poles with the positive chirality and the same number of ones with the negative chirality for both toroidal and poloidal directions in all surfaces, this phase belongs to weak TI.

While one can readily obtain correspondences between these kinds of pole winding number structures and possible surface bands' topologies of general 3D insulators, we leave the general classification of 3D BBCs for upcoming future work. This classification will of course also depend on the underlying lattice symmetries, as weak insulates can only be viewed as a stacking of 2D quantum spin Hall systems when the space group allows for it \cite{Slager2013}.

\section{Conclusions}

While the bulk-boundary correspondence is a hallmark in  the analysis of topological matter, it has never been constructed on general grounds. Indeed, there already exist counterexamples for, e.g., the 1D Berry phase's bulk-boundary correspondence even though this entails a noninteracting system.
In addition, one should consider different kinds of topological invariants for each specific BBC under consideration, which depends on the underlying symmetries of the symmetry protected topological (SPT) phase. Finally, there are also practical issues, like gauge fixing problems, in terms of direct numerical computation. 

We have here  developed a new way to extract the profiles of in-gap modes from bulk wave functions in a unified manner for each dimension.
In 1D, we have obtained bulk numbers for the exact and even-odd predictions of in-gap boundary modes when the bulk Hamiltonian has boundaries such as open boundaries, and junctions.
This is done by connecting the bulk Hamiltonian to the one with boundaries continuously via a control parameter $\beta$.
As $\beta$ increases, a number of modes come out of the bulk band continua into the gap.  We have shown that the net number of such in-gap modes is determined by considering the pole determinant in the vicinity of the valence and conduction band edges.
From this, we can derive explicit formulae for the numbers for the bulk-boundary correspondence.
Since varying $\beta$ requires a tiny numerical cost, one can accordingly calculate those bulk numbers with a computational expense that is similar to that of a Berry phase or a Wilson loop calculation\cite{Alexandradinata2014,Taherinejad2014,Alexandradinata2016}, provided that the inter-orbital overlaps are short-ranged.
Furthermore, there is no gauge fixing problem in evaluating our bulk numbers since they are obtained from the pole structure of the Green's function while we have this problem in calculating many known topological invariants, and it has been a crucial issue to resolve this obstacle\cite{Fu2007,Fukui2007,Soluyanov2011,Yu2011,Abanin2013,Lau2016}.

More specifically, we have also shown that those numbers work well in various concrete models.  In particular, our numbers even hold in case of  `counterexamples', in which  the Berry phase's bulk-boundary correspondence does not hold, such as the incommensurate SSH model and the double SSH model.
Moreover, we showed that our bulk numbers can be successfully applied in the experimentally motivated study of the bulk-boundary correspondence of graphene for various kinds of boundaries, such as the open boundaries along arbitrary directions and grain boundaries.

In 2D, on the other hand, one is mainly interested in how many surface bands connect the valence and conduction bands.
The metallicity of those kind of surface bands is namely topologically protected against symmetry-protecting perturbations.
We have shown that topological structure of the surface bands is naturally encoded in the pole's evolutions along the the valence or conduction band portals during the effective periodic process that shifts one of the edges by a unit cell while leaving the electronic structure in the gap invariant.
We have defined a different kind of the winding number, the pole winding number, and according chirality for each pole that can predict the topological characters of each surface band in the gap.
These invariants also can be applied to the bulk-boundary correspondence for arbitrary 2D insulators regardless of their underlying symmetries.
Finally, in 3D, one can extract profiles of surface bands by evaluating the pole winding numbers in all the possible sub-BZs.
While one then has more diverse types of surface bands than in the 2D case, we have shown that 3D TBIs, for example, naturally fit within this scheme.
Although it is clear that the full classification of all possible surface band types find a natural correspondence to the pole winding numbers, we will pursuit the explicit evaluation in upcoming future work. 
In this regard it would also be interesting to find a connection to the found quantities in the context of Floquet system, where similar BBC evaluations have been performed recently.

\acknowledgements
This work was supported by the ERC Starting Grant No. 679722 and the Research Center Program of the Institute for Basic Science in Korea No. IBS-R009-D1. J.-W. Rhim thanks G. Y. Cho for useful discussions.

\appendix

\newpage

\section{$N_\mathcal{V} =2 $ case}\label{app:N_V=2}

When $N_\mathcal{V} = 2$, one can derive a simple formula for the pole-determinant $A_\beta(\varepsilon)$ at $\beta=1$ near the band edge as follows.
The pole-matrix for this case is given by
\begin{widetext}
\begin{align}
\mathcal{A}_1(\varepsilon^\alpha) \approx \frac{(-1)^{\alpha}}{\delta\varepsilon} \bpm \mathcal{D}^{(1)}_{\alpha}|_{1,1} & \mathcal{D}^{(1)}_{\alpha}|_{1,2} \\ \mathcal{D}^{(1)}_{\alpha}|_{2,1} & \mathcal{D}^{(1)}_{\alpha}|_{2,2}\epm + \bpm 1-\mathcal{D}^{(2)}_{\alpha}|_{1,1} & -\mathcal{D}^{(2)}_{\alpha}|_{1,2} \\ -\mathcal{D}^{(2)}_{\alpha}|_{1,1} & 1-\mathcal{D}^{(2)}_{\alpha}|_{1,2} \epm \label{eq:A_mat_2by2_0}
\end{align}
\end{widetext}
where $\mathcal{D}^{(q)}_{\alpha}|_{i,j}$ is the $i,j$-th element of $\mathcal{D}^{(q)}_{\alpha}$.
Recall that $\varepsilon^c = \epsilon^\mathrm{min} - \delta\varepsilon$ and $\varepsilon^v = \epsilon^\mathrm{max} + \delta\varepsilon$ where $\epsilon^{\mathrm{min}}$ and $\epsilon^{\mathrm{max}}$ are the energies of the conduction band minima and the valence band maxima, and $\delta\varepsilon$ is an infinitesimal positive number.
We consider the two cases, $\det\mathcal{D}^{(1)}_{\alpha} = 0$ and $\det\mathcal{D}^{(1)}_{\alpha} \neq 0$, separately.
First, when $\det\mathcal{D}^{(1)}_{\alpha} = 0$, the pole-determinant is written as
\begin{widetext}
\begin{align}
A_1(\varepsilon^\alpha) \approx & \frac{m^\alpha}{\delta\varepsilon}\left( \mathcal{D}^{(1)}_{\alpha}|_{1,1} + \mathcal{D}^{(1)}_{\alpha}|_{2,2} - \mathcal{D}^{(2)}_{\alpha}|_{2,2}\mathcal{D}^{(1)}_{\alpha}|_{1,1} + \mathcal{D}^{(2)}_{\alpha}|_{2,1}\mathcal{D}^{(1)}_{\alpha}|_{1,2} - \mathcal{D}^{(2)}_{\alpha}|_{1,1}\mathcal{D}^{(1)}_{\alpha}|_{2,2} + \mathcal{D}^{(2)}_{\alpha}|_{1,2}\mathcal{D}^{(1)}_{\alpha}|_{2,1}\right). \label{eq:A_mat_2by2}%\nonumber\\
\end{align}
\end{widetext}
From (\ref{eq:D1}) and (\ref{eq:D2}), we have
\begin{widetext}
\begin{align}
\mathcal{D}^{(2)}_{\alpha}|_{2,2}\mathcal{D}^{(1)}_{\alpha}|_{1,1} - \mathcal{D}^{(2)}_{\alpha}|_{2,1}\mathcal{D}^{(1)}_{\alpha}|_{1,2} =& \sideset{}{'}\sum_{n^\prime,k^\prime}\sum_{l=1}^{N^*} \frac{ \langle v_2 | n^\prime,k^\prime \rangle \langle n^\prime,k^\prime | \mathcal{V}_b|v_2 \rangle \langle v_1 | \alpha,k^*_l \rangle \langle \alpha,k^*_l | \mathcal{V}_b|v_1 \rangle }{\epsilon^* - \epsilon_{n^\prime,k^\prime}} \nonumber \\
&-\sideset{}{'}\sum_{n^\prime,k^\prime}\sum_{l=1}^{N^*} \frac{ \langle v_2 | n^\prime,k^\prime \rangle \langle n^\prime,k^\prime | \mathcal{V}_b|v_1 \rangle \langle v_1 | \alpha,k^*_l \rangle \langle \alpha,k^*_l | \mathcal{V}_b|v_2 \rangle }{\epsilon^* - \epsilon_{n^\prime,k^\prime}}, \label{eq:A3} \\
\mathcal{D}^{(2)}_{\alpha}|_{1,1}\mathcal{D}^{(1)}_{\alpha}|_{2,2} - \mathcal{D}^{(2)}_{\alpha}|_{1,2}\mathcal{D}^{(1)}_{\alpha}|_{2,1} =& \sideset{}{'}\sum_{n^\prime,k^\prime}\sum_{l=1}^{N^*} \frac{ \langle v_1 | n^\prime,k^\prime \rangle \langle n^\prime,k^\prime | \mathcal{V}_b|v_1 \rangle \langle v_2 | \alpha,k^*_l \rangle \langle \alpha,k^*_l | \mathcal{V}_b|v_2 \rangle }{\epsilon^* - \epsilon_{n^\prime,k^\prime}} \nonumber \\
&-\sideset{}{'}\sum_{n^\prime,k^\prime}\sum_{l=1}^{N^*} \frac{ \langle v_1 | n^\prime,k^\prime \rangle \langle n^\prime,k^\prime | \mathcal{V}_b|v_2 \rangle \langle v_2 | \alpha,k^*_l \rangle \langle \alpha,k^*_l | \mathcal{V}_b|v_1 \rangle }{\epsilon^* - \epsilon_{n^\prime,k^\prime}} \label{eq:A4}
\end{align} 
\end{widetext}
for the last four terms in (\ref{eq:A_mat_2by2}), and
\begin{align}
\sum_{i=1}^2\mathcal{D}^{(1)}_{\alpha}|_{i,i} =& \sum_{i=1}^2\langle v_i | \sum_{l=1}^{N^*} |\alpha,k^*_l\rangle \langle \alpha,k^*_l | \mathcal{V}_b | v_j \rangle \\
=& \sum_{l=1}^{N^*} \langle \alpha,k^*_l | \mathcal{V}_b |\alpha,k^*_l\rangle \\
\equiv & \langle \mathcal{V}_b\rangle^* \label{eq:Vstar} \\
\nonumber \\
\nonumber
\end{align}
for the first two terms (\ref{eq:A_mat_2by2}).
Taking into account the identity, 
\begin{align}
C_1 =& \langle n,k | \mathcal{V}_b| n,k \rangle \langle \alpha,k^*_l | \mathcal{V}_b| \alpha,k^*_l \rangle \nonumber \\
& -\langle n,k | \mathcal{V}_b| \alpha,k^*_l \rangle \langle \alpha,k^*_l | \mathcal{V}_b| n,k \rangle \\
=& \langle n,k | \mathcal{V}_b| v_1 \rangle \langle v_1 | n,k \rangle \langle \alpha,k^*_l| \mathcal{V}_b| v_2 \rangle \langle v_2 | \alpha,k^*_l \rangle \nonumber \\
&+ \langle n,k | \mathcal{V}_b| v_2 \rangle \langle v_2 | n,k \rangle \langle \alpha,k^*_l | \mathcal{V}_b| v_1 \rangle \langle v_1 | \alpha,k^*_l \rangle \nonumber \\
& \langle n,k | \mathcal{V}_b| v_1 \rangle \langle v_1 | \alpha,k^*_l \rangle \langle \alpha,k^*_l | \mathcal{V}_b| v_2 \rangle \langle v_2 | n,k \rangle \nonumber \\
&- \langle n,k | \mathcal{V}_b| v_2 \rangle \langle v_2 | \alpha,k^*_l \rangle \langle \alpha,k^*_l | \mathcal{V}_b| v_1 \rangle \langle v_1 | n,k \rangle,
\end{align}
we have
\begin{widetext}
\begin{align}
(\mathrm{\ref{eq:A3}}) + (\mathrm{\ref{eq:A4}}) =& \sideset{}{'}\sum_{n^\prime,k^\prime}\sum_{l=1}^{N^*} \frac{\langle n^\prime,k^\prime | \mathcal{V}_b| n^\prime,k^\prime \rangle \langle \alpha,k^*_l  | \mathcal{V}_b| \alpha,k^*_l \rangle - \langle n^\prime,k^\prime | \mathcal{V}_b| \alpha,k^*_l  \rangle \langle \alpha,k^*_l  | \mathcal{V}_b| n^\prime,k^\prime \rangle}{\epsilon^* - \epsilon_{n^\prime,k^\prime}} \\
\equiv & \sideset{}{'}\sum_{n^\prime,k^\prime}\sum_{l=1}^{N^*} \frac{d^{n^\prime,k^\prime}_{\alpha,k^*_l }}{\epsilon^* - \epsilon_{n^\prime,k^\prime}}. \label{eq:dstar}
\end{align}
\end{widetext}
Substituting (\ref{eq:Vstar}) and (\ref{eq:dstar}) into (\ref{eq:A_mat_2by2}), we obtain
\begin{align}
A_1(\varepsilon^\alpha) \approx \frac{m^\alpha}{\delta\varepsilon} \left( \langle \mathcal{V}_b\rangle^* - \sideset{}{'}\sum_{n^\prime,k^\prime}\sum_{l=1}^{N^*} \frac{ d^{n^\prime,k^\prime}_{\alpha, k^*}}{\epsilon^* - \epsilon_{n,k}} \right)
\end{align}
for $\det\mathcal{D}^{(1)}_{n,\alpha} = 0$.

On the other hand, when $\det\mathcal{D}^{(1)}_{\alpha} \neq 0$, only the first term of (\ref{eq:A_mat_2by2_0}) is important.
In this case, the pole-determinant becomes
\begin{align}
A_1(\varepsilon^\alpha) \approx \frac{1}{\delta\varepsilon^2} \mathrm{det}\mathcal{D}^{(1)}_{\alpha}.
\end{align}

From (\ref{eq:D1}), we have
\begin{widetext}
\begin{align}
\mathrm{det}\mathcal{D}^{(1)}_{\alpha} =& \langle v_1 | \sum_{l_1=1}^{N^*} |\alpha,k^*_{l_1}\rangle \langle \alpha,k^*_{l_1} | \mathcal{V}_b | v_1 \rangle \langle v_2 | \sum_{l_2=1}^{N^*} |\alpha,k^*_{l_2}\rangle \langle \alpha,k^*_{l_2} | \mathcal{V}_b | v_2 \rangle \nonumber \\
& - \langle v_1 | \sum_{l_1=1}^{N^*} |\alpha,k^*_{l_1}\rangle \langle \alpha,k^*_{l_1} | \mathcal{V}_b | v_2 \rangle \langle v_2 | 
\sum_{l_2=1}^{N^*} |\alpha,k^*_{l_2}\rangle \langle \alpha,k^*_{l_2} | \mathcal{V}_b | v_1 \rangle \\
=& \frac{1}{2}\sum_{l_1=1}^{N^*}\sum_{l_2=1}^{N^*} \langle \alpha,k^*_{l_1} | \mathcal{V}_b \sum_{i=1}^2 | v_i \rangle \langle v_i |\alpha,k^*_{l_1}1\rangle \langle \alpha,k^*_{l_2} | \mathcal{V}_b \sum_{j=1}^2 | v_j \rangle \langle v_j |\alpha,k^*_{l_2}\rangle \nonumber \\
&-\frac{1}{2}\sum_{l_1=1}^{N^*}\sum_{l_2=1}^{N^*} \langle \alpha,k^*_{l_1} | \mathcal{V}_b \sum_{i=1}^2 | v_i \rangle \langle v_i |\alpha,k^*_{l_2}\rangle \langle \alpha,k^*_{l_2} | \mathcal{V}_b \sum_{j=1}^2 | v_j \rangle \langle v_j |\alpha,k^*_{l_1}\rangle \\
=& \frac{1}{2}\sum_{l_1=1}^{N^*}\sum_{l_2=1}^{N^*} d^{\alpha,k^*_{l_1}}_{\alpha,k^*_{l_2}}
\end{align}
\end{widetext}
which leads to
\begin{align}
A_1(\varepsilon^\alpha) \approx \frac{1}{2\delta\varepsilon^2}\sum_{l_1}\sum_{l_2} d^{\alpha, k^*_{l_1}}_{\alpha, k^*_{l_2}}.
\end{align}

\section{Rice-Mele model}\label{app:rice-mele}

The eigenenergy and eigenvectors of Rice-Mele Hamiltonian (\ref{eq:RM_ham}) are given by
\begin{align}
\epsilon_{n,k} = (-1)^n \sqrt{s(k)^2 + \Delta^2}
\end{align}
and
\begin{align}
|1\rangle &= \bpm v^A_{1,k} \\ v^B_{1,k} \epm = \bpm -\sqrt{\frac{1}{2}+\frac{\Delta}{2\epsilon_{1,k}}} \\ \sqrt{\frac{1}{2}+\frac{\Delta}{2\epsilon_{2,k}}}e^{i\phi_k} \epm, \label{eq:rm_eigvec1} \\
|2\rangle &= \bpm v^A_{2,k} \\ v^B_{2,k} \epm =\bpm \sqrt{\frac{1}{2}+\frac{\Delta}{2\epsilon_{2,k}}} \\ \sqrt{\frac{1}{2}+\frac{\Delta}{2\epsilon_{1,k}}}e^{i\phi_k} \epm \label{eq:rm_eigvec2}
\end{align}
where the first and second rows represent A and B sites each.
Here, $s(k) =\sqrt{2(t^2+\delta^2) + 2(t^2-\delta^2)\cos k}$ and $e^{i\phi_k} = -2 (t\cos k/2 - i\delta\sin k/2)/s(k)$.
The corresponding Bloch wave functions are represented as
\begin{align}
|n,k\rangle = \frac{1}{\sqrt{N}}\sum_{i} \left( v^A_{n,k}e^{ikx_{A,i}} |a_i\rangle + v^B_{n,k}e^{ikx_{B,i}} |b_i\rangle \right)
\end{align}
where $x_{A,i}$ and $x_{B,i}$ are the positions of the A and B sites in the $i$-th unit cell, and $N\rightarrow\infty$ is the total number of unit cells.
Here, $|a_i\rangle$ and $|b_i\rangle$ are the local orbitals at A and B sites in the $i$-th unit cell.
Also, note that the valence band shows its maximum at $k=\pi$ when $|t|>|\delta|$, and at $k=0$ when $|t|<|\delta|$.

To make a commensurate termination, we apply a local operator which cancels out all the hoppings across the boundary between the first and second unit cells.
This operator is given by
\begin{align}
\mathcal{V}_b = (t-\delta)\left( a^\dag_2 b_1 + b^\dag_1 a_2 \right)
\end{align}
which yields the matrix element
\begin{align}
\langle n,k | \mathcal{V}_b | n^\prime, k^\prime \rangle =& \frac{t-\delta}{N} \Big( v^{A*}_{n,k} v^{B}_{n^\prime,k^\prime} e^{-ikx_{A,2}}e^{ik^\prime x_{B,1}}  \nonumber\\
& + v^{B*}_{n,k} v^{A}_{n^\prime,k^\prime} e^{-ikx_{B,1}}e^{ik^\prime x_{A,2}} \Big). \label{eq:RM_Vmat}
\end{align}

Let us calculate $\langle \mathcal{V}_b\rangle^*$.
Here, we only consider the $|t|>|\delta|$ case since we have the same result for $|t|<|\delta|$ case.
In this case, the band edge is located at $k^* = \pi$ with energies $\epsilon^*_v = \epsilon_{1,\pi} = -\sqrt{4\delta^2 + \Delta^2}$ and $\epsilon^*_c = \epsilon_{2,\pi} = \sqrt{4\delta^2 + \Delta^2}$ for the valence and conduction bands each.
Then, we have
\begin{align}
\langle \mathcal{V}_b\rangle^* =& \langle 1,k^* | \mathcal{V}_b | 1, k^* \rangle \\
=& -2\frac{t-\delta}{N}\frac{\delta}{|\delta|} \sqrt{\frac{1}{2}+\frac{\Delta}{2\epsilon_{1,k}}} \sqrt{\frac{1}{2}+\frac{\Delta}{2\epsilon_{2,k}}} \\
=& -\frac{2}{N} \frac{\delta(t-\delta)}{\sqrt{4\delta^2+\Delta^2}}
\end{align}
for the valence band, and
\begin{align}
\langle \mathcal{V}_b\rangle^* = \langle 2,k^* | \mathcal{V}_b | 2, k^* \rangle= \frac{2}{N} \frac{\delta(t-\delta)}{\sqrt{4\delta^2+\Delta^2}}
\end{align}
for the conduction band because $e^{i\phi_\pi} = i\delta/|\delta|$ and $e^{ik^*(x_{B,1}-x_{A,2})} = -i$.

Now, we evaluate (\ref{eq:dstar}) for $|t|>|\delta|$ case.
From (\ref{eq:dstar}) and (\ref{eq:RM_Vmat}), the denominator of the summand of (\ref{eq:dstar}) is evaluated as
\begin{widetext}
\begin{align}
d^{n,k}_{m,k^*} = (-1)^{m+1}\frac{(t-\delta)^2}{N^2} \left\{  -\frac{1}{2} - \frac{(-1)^n\Delta^2}{2\epsilon_{1,k} \epsilon_{1,k^*}} - \frac{(-1)^n}{2}\sqrt{1-\frac{\Delta^2}{\epsilon_{1,k}^2}}\sqrt{1-\frac{\Delta^2}{\epsilon_{1,k^*}^2}}\cos\left( \phi_k - \frac{k}{2} +\frac{\pi}{2}\left(1 - \frac{\delta}{|\delta|}\right) \right) \right\}
\end{align}
\end{widetext}
where $m$ and $n$ are the band indices.
From this, we obtain
\begin{widetext}
\begin{align}
\sideset{}{'}\sum_{n,k} \frac{d^{n,k}_{m,k^*}}{\epsilon^* - \epsilon_{n,k}} =& \sideset{}{'}\sum_{k} (-1)^{m+1}\frac{(t-\delta)^2}{N^2} \frac{\Delta^2 - \epsilon_{1,k^*}^2 + \sqrt{\epsilon_{1,k}^2 -\Delta^2 }\sqrt{\epsilon_{1,k^*}^2 -\Delta^2 }\cos\left( \phi_k - \frac{k}{2} + \frac{\pi}{2}\left(1 - \frac{\delta}{|\delta|}\right) \right) }{ \epsilon_{1,k^*} \left( \epsilon_{1,k^*}^2 -\epsilon_{1,k}^2  \right) } \\
=& \sideset{}{'}\sum_{k} (-1)^{m+1}\frac{(t-\delta)^2}{N^2}\frac{-s(k^*)^2 + \frac{\delta}{|\delta|}s(k^*)\sqrt{\epsilon_{1,k}^2 -\Delta^2 }\left( \cos \phi_k \cos\frac{k}{2} + \sin\phi_k\sin\frac{k}{2} \right) }{ \epsilon_{1,k^*} \left( \epsilon_{1,k^*}^2 -\epsilon_{1,k}^2  \right) } \\
=& \sideset{}{'}\sum_{k} (-1)^{m+1}\frac{(t-\delta)^2}{N^2}\frac{-s(k^*)^2 + \frac{\delta}{|\delta|} s(k^*)\left( -2t \cos^2\frac{k}{2} + 2\delta \sin^2\frac{k}{2} \right) }{ \epsilon_{1,k^*} \left( \epsilon_{1,k^*}^2 -\epsilon_{1,k}^2  \right)} \\
=& \frac{(-1)^{m}}{N}\frac{\delta(t-\delta)}{\sqrt{4\delta^2+\Delta^2}} \label{eq:RM_final}
\end{align}
\end{widetext}
because $s(k^*) = 2|\delta|$ when $|t|>|\delta|$.
As a result, the pole-determinant becomes
\begin{align}
A_1(\varepsilon^c) = A_1(\varepsilon^v) = \frac{1}{\delta\varepsilon N}  \frac{\delta(t-\delta)}{\sqrt{4\delta^2+\Delta^2}}
\end{align}
where we applied $m=1$ around the valence band edge ($A_1(\varepsilon^v)$), and $m=2$ around the valence band edge ($A_1(\varepsilon^c)$) in (\ref{eq:RM_final}).
Finally we obtain
\begin{align}
\frac{A_1(\varepsilon^\alpha)}{|A_1(\varepsilon^\alpha)|} = \mathrm{sgn}(t\delta)
\end{align}
because $|t|>|\delta|$.

\begin{figure}
	\begin{center}
		\includegraphics[width=\columnwidth]{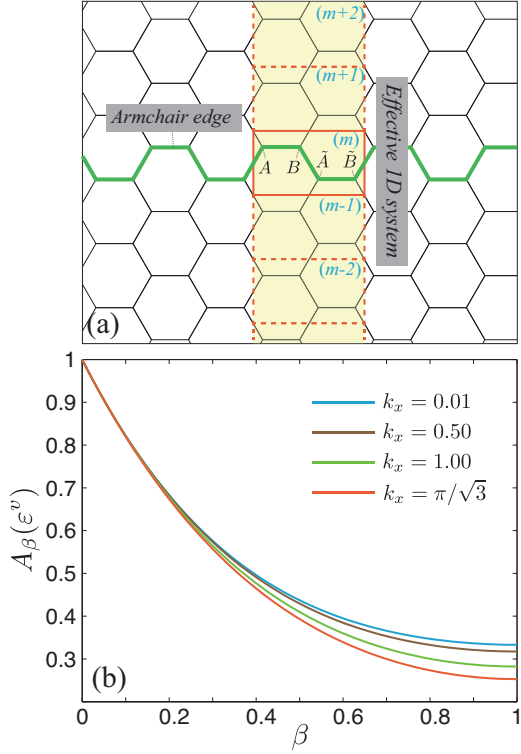}
	\end{center}
	\caption{(a) The band structure of the finite Double SSH chain with 1000 unit cells as a function of $t^\prime/t$, where $t^\prime$ is the inter-chain coupling. Here, $\delta/t = 0.2$, and $\epsilon_0/t = 0.3$.}
	\label{fig:agnr}
\end{figure}

\section{Armchair graphene nanoribbon}\label{app:agnr}

To deal with the armchair GNR, let us consider the four-site unit cell for graphene as presented in Fig.
Within the basis $c_\mathbf{k} = \bpm  c_{A,\mathbf{k}} & c_{B,\mathbf{k}} & c_{\tilde{A},\mathbf{k}} & c_{\tilde{B},\mathbf{k}}\epm^\mathrm{T}$, the Hamiltonian's matrix is given by
\begin{align}
H_{\mathrm{AGNR}} = \bpm 0 & f^*_1(\mathbf{k}) & 0 & f_2(\mathbf{k}) \\ f_1(\mathbf{k}) & 0 & f^*_2(\mathbf{k}) & 0 \\ 0 & f_2(\mathbf{k}) & 0 & f^*_1(\mathbf{k}) \\ f^*_2(\mathbf{k}) & 0 & f_1(\mathbf{k}) & 0 \epm
\end{align}
where $f_1(\mathbf{k}) = -e^{iak_x/\sqrt{3}}$, and $f_2(\mathbf{k}) = -2\cos (ak_y/2) e^{iak_x/2\sqrt{3}}$.
Let us assume $a=1$.
The energy spectrum of $H_{\mathrm{AGNR}}$ is given by
\begin{align}
\epsilon^{\eta_1,\eta_2}_{k_x,k_y} = \eta_1\sqrt{3+2\cos k_y+4\eta_2\cos\frac{\sqrt{3}k_y}{2}\cos\frac{k_x}{2}}
\end{align}
where $\eta_i = \pm$.

If we fix $k_x$, $H_{\mathrm{AGNR}}$ is the Hamiltonian describing the effective 1D system along $y$ direction with hopping parameters $-e^{\pm ik_x/\sqrt{3}}$ and $-e^{\pm ik_x/2\sqrt{3}}$.
The local operator which makes edges between two neighboring unit cells is given by
\begin{align}
\mathcal{V}_b = \bpm 0 & 0 & 0 & e^{i\frac{k_x}{2\sqrt{3}}} \\ 0 & 0 & e^{-i\frac{k_x}{2\sqrt{3}}} & 0 \\ 0 & e^{i\frac{k_x}{2\sqrt{3}}} & 0 & 0 \\ e^{-i\frac{k_x}{2\sqrt{3}}} & 0 & 0 & 0 \epm
\end{align}
in the basis $c_\mathcal{V} = \bpm  c_{m,A} & c_{m,B} & c_{m+1,\tilde{A}} & c_{m+1,\tilde{B}} \epm^\mathrm{T}$ where $m$ is the index of the unit cell of the effective 1D system.

From these, the pole-matrix is evaluated as
\begin{align}
\mathcal{A}_\beta(\varepsilon)|_{ij} = \delta_{ij} - \frac{\beta}{2\pi}\sum_{\eta_1,\eta_2}\int_{-\pi}^\pi dk_y \frac{\langle i | \epsilon^{\eta_1,\eta_2}_{k_x,k_y} \rangle \langle \epsilon^{\eta_1,\eta_2}_{k_x,k_y} | \mathcal{V}_b | j \rangle }{\varepsilon - \epsilon^{\eta_1,\eta_2}_{k_x,k_y}}
\end{align}
where $i$ and $j$ run over all four elements of $c_\mathcal{V}$.
Note that $| \epsilon^{\eta_1,\eta_2}_{k_x,k_y} \rangle$ is the real-space form of the Bloch wave function, not just the eigenvector of $H_{\mathrm{AGNR}}$.

While we are interested in the behavior of the pole-determinant in the vicinity of the conduction and valence band edges, $\varepsilon^\alpha = -m^\alpha (|\sin\sqrt{3}k_x/2| - \delta\varepsilon)$, we have $A_\beta(\varepsilon^c) = A_\beta(\varepsilon^v)$ due to chiral symmetry.
Here, $m^{c(v)}=-1(1)$.
Therefore, we only plot $A_\beta(\varepsilon^v)$ as a function of $\beta$ for various values of $k_x$ in Fig.~\ref{fig:agnr}(b).
We accordingly find that the pole-determinant of the armchair GNR is always positive in $0 \leq \beta \leq 1$ at any $k_x$, which means both $M$, $P$, and $P_\mathrm{half}$ are zero, and there is no in-gap mode.

\section{Kane-Mele model}\label{app:pole-winding}

In this section, we revisit the Kane-Mele model\cite{Kane2005} briefly. 
The Kane-Mele model is given by
\begin{align}
\mathcal{H}_\mathrm{KM} =& t\sum_{\langle ij\rangle} c^\dag_i c_j + i\lambda_\mathrm{SO}\sum_{\langle\langle ij\rangle\rangle}\nu_{ij}c^\dag_i s^z c_j + \lambda_v\sum_i\xi_i c^\dag_i c_i \nonumber\\
& + i\lambda_\mathrm{R}\sum_{\langle ij\rangle}c^\dag_i (\mathbf{s}\times\hat{\mathbf{d}}_{ij}) c_j 
\end{align}
where the first term is the nearest neighboring hopping process of graphene, the second term is the intrinsic spin-orbit coupling between the next neighboring sites, the third term is the onsite mass term breaks the sublattice symmetry, and the final term is the Rashba spin-orbit coupling\cite{Kane2005}.
$\nu_{ij}=1(-1)$ if the path from the $j$-th site to the $i$-th site through two bonds is counterclockwise(clockwise).
$s^z$ is the Pauli matrix for spin degrees of freedom, and $\hat{\mathbf{d}}_{ij}$ is the unit vector for the direction from the $j$-th site to the $i$-th site.
Note that we omit labels for spins.

By performing a partial Fourier transformation along the $y$ direction, we obtain an 1D effective Hamiltonian for fixed $k_y$. This is illustrated in Fig.~\ref{fig:winding_km}(a) in the main text.
We then make the open boundary by setting $\beta_1=1$.
As shown in Fig.~\ref{fig:pole_evolution}, we may accordingly determine the evolutions of the surface bands as a function of $\beta_2$, which controls the magnitude of the onsite potential $\mathcal{V}_2$ for the orbitals in the left end unit cell. Similarly, this also conveys how these lead to the pole winding on the valence band portal drawn by red dashed curves.
Here, we assume that $V_0 = -1$ so that the left-localized states are going down to the lower energies for increasing $\beta_2$.
We compare two cases, one for the topologically nontrivial phase with $\lambda_v = 0.1t$, $\lambda_\mathrm{SO} = -0.06t$, and $\lambda_\mathrm{R} = 0.05t$, and the other for the trivial insulating phase with $\lambda_v = 0.4t$, $\lambda_\mathrm{SO} = -0.06t$, and $\lambda_\mathrm{R} = 0.05t$.
The nontrivial phase hosts the quantum spin Hall effect.

\bibliography{bbc}

\begin{thebibliography}{79}
\bibitem [{not()}]{note1}%
  \BibitemOpen
  \href@noop {} {}\bibinfo {note} {{Here, we consider the weak TI with
  infinitesimal interlayer coupling for the simplicity. However, the conclusion
  holds in general even though the bands are weakly dispersive along the
  stacking direction.}}\BibitemShut {Stop}%
\end{thebibliography}

\end{document}